\documentclass[prd, preprint, tightenlines, a4paper, eqsecnum, nofootinbib]{revtex4-2}
\usepackage[T1]{fontenc}
\usepackage{physics}
\usepackage{braket}
\usepackage{amsmath}
\usepackage{mathtools}
\usepackage{amssymb}
\usepackage{stackrel}
\usepackage{mathrsfs}
\usepackage{bigints}
\usepackage{bbm} 
\usepackage{graphicx}
\usepackage{float}
\usepackage{mdframed}
\usepackage{makecell} 
\usepackage{hhline}
\usepackage[usernames,dvipsnames,svgnames,table]{xcolor}
\usepackage{multirow}
\usepackage{tabularx}
\usepackage{graphbox} 
\usepackage{url}
\usepackage{hyperref}

\DeclareMathOperator{\Imag}{Im}
\DeclareMathOperator{\Real}{Re}
\newcommand{\mini}{\scriptscriptstyle}
\makeatletter
\def\l@subsubsection#1#2{}
\makeatother
\begin{document}

\title{On the role of boundary conditions \\ within Ho\u{r}ava-Lifshitz gravity}

\author{Lissa de Souza Campos$^{1,2}$}
\email{lissa.desouzacampos01@universitadipavia.it}
\author{Claudio Dappiaggi$^{1,2}$}
\email{claudio.dappiaggi@unipv.it}
\author{Denis Sina$^1$}
\email{denis.sina01@universitadipavia.it}
\affiliation{$^1$Dipartimento di Fisica, Universit\`a degli Studi di Pavia, Via Bassi, 6, 27100 Pavia, Italy}
\affiliation{$^2$Istituto Nazionale di Fisica Nucleare -- Sezione di Pavia, Via Bassi, 6, 27100 Pavia, Italy}

\date{\today}
\vspace{1cm}
\begin{abstract}
  On a class of four-dimensional Lifshitz spacetimes with critical exponent $z=2$, including a hyperbolic and a spherical Lifshitz topological black hole, we consider a real Klein-Gordon field. Using a mode-decomposition, we split the equation of motion into a radial and into an angular component. As first step, we discuss under which conditions on the underlying parameters, we can impose to the radial equation boundary conditions of Robin type and whether bound state solutions do occur. Subsequently, we show that, whenever bound states are absent, one can associate to each admissible boundary condition a ground and a KMS state whose associated two-point correlation function is of local Hadamard form. \\


\end{abstract}

{
\let\clearpage\relax
\maketitle
}
\newpage
\tableofcontents
\newpage


\section{Introduction}

In the quest of finding a quantization scheme for the gravitational field, Ho\u{r}ava-Lifshitz gravity,   \cite{Horava:2009uw}, was first introduced
as a theory in which the underlying spacetime possesses a time coordinate
$t$ and spatial counterparts $\underline{x}$ appearing with a different scaling behaviour
\begin{equation}
	\label{eq:scaling lifshitz}
	t\mapsto \delta^z t \quad\text{ and }\quad \underline{x}\mapsto\delta\underline{x}  \quad\text{ with }\quad z>1\;\;\textrm{and}\;\;\delta>0.
\end{equation}
While we will not focus on the developments towards a quantum theory of gravitation, referring an interested reader to the recent review \cite{Wang:2017brl}, we stress that also different applications of these models have been discussed ranging from cosmology, to quantum critical systems \cite{Hartnoll:2009sz}, to condensed matter physics, see {\em e.g.} \cite{Horava:2011gd} and references therein, and to the AdS/CFT correspondence.

Especially motivated from this last framework, a particular interest has been devoted to studying the behaviour both at a classical and at a quantum level of a scalar field living either on a Lifshitz spacetime or on asymptotically Lifshitz black holes \cite{Kachru:2008yh,Giacomini:2012hg,Quinta:2016eql}. A remarkable property of all these manifolds lies in the fact that the underlying metric possesses a singular behaviour along a spatial direction and therefore the spacetime is not globally hyperbolic. As
a consequence, the dynamics of a scalar field cannot be fully determined only by assigning initial data, but one needs also to specify an asymptotic behaviour towards the singularity of the metric, exactly as it occurs when one considers Lorentzian manifolds with a timelike boundary, such as
asymptotically anti-de Sitter spacetimes. A close scrutiny of the literature unveils that, in the case of backgrounds of Lifshitz type, only Dirichlet boundary conditions have been considered.

Yet, in the past few years, starting from the work of Ishibashi and Wald \cite{Ishibashi:2003jd, Ishibashi:2004wx} on the dynamics of free Bosonic
fields on anti-de Sitter spacetimes, it has become clear that one can consider a much larger class of boundary conditions of Robin type, yielding a well-defined dynamics and, in addition, admitting an associated, full-fledged quantization scheme. This enlarged set of options has been studied
in detail on anti-de Sitter spacetime \cite{Dappiaggi:2016fwc,Dappiaggi:2017wvj,Dappiaggi:2018xvw}, on a rotating BTZ black hole \cite{Bussola:2017wki,Bussola:2018iqj} and recently on massless hyperbolic black holes \cite{Campos:2020lpt}.

Inspired by these works we consider a massive, real scalar field, with an
arbitrary coupling to scalar curvature and we investigate whether one can
extend the class of admissible boundary conditions on a four dimensional Lifshitz spacetime with dynamical critical exponent $z=2$ and on its generalizations to a hyperbolic and to a spherical Lifshitz topological black hole, as determined in \cite{Mann:2009yx}. Without entering here into the technical details of the analysis, our investigation leads to two main results. On the one hand, we prove that, in all these spacetimes, there
exists a specific range of the underlying parameters, namely the mass and
the coupling to scalar curvature, for which a general class of boundary conditions of Robin type is admissible. In this respect, it is worth emphasizing that our result is compatible with \cite{Keeler:2012mb}. On the other hand, we show that, similarly to what happens on anti-de Sitter spacetimes, one can stumble into the so-called bound state modes. More precisely, since all the spacetimes that we consider are static, it is possible to study the dynamics of the underlying field and the implementation of the boundary conditions by considering the Fourier transform along the time coordinate $t$. As a consequence one can show that, depending on the range of boundary conditions considered, $\omega$, the Fourier parameter associated to $t$, is not necessarily real but it can also take value in a discrete set of purely imaginary frequencies, the above mentioned bound states modes. Here we confirm and extend the results of \cite{Andrade:2012xy}, in which it was shown the occurrence of instabilities for a scalar field on Lifshitz spacetime for a specific range of the field parameters.

Although at a classical level, this feature is not particularly problematic, it has more severe consequences at a quantum level since it entails the existence of modes, exponentially growing in time. As a consequence, whenever such bound states do appear, it is not possible to construct the two-point function of a ground state associated to the underlying scalar field, as already observed in \cite{Dappiaggi:2016fwc}.

After having established in which range of boundary conditions, bound states do not occur, in the remaining part of the paper we focus our attention on all other cases and we show how to construct for each of them a two-point correlation function both for a ground state and for a KMS/thermal
state at arbitrary temperature. It is important to observe that all these
two-point functions obey to the underlying equation of motion, they implement the canonical commutation relations as well as the chosen boundary conditions and, in addition they are of local Hadamard form. Such property
guarantees the possibility of constructing Wick-ordered observables which
is an important prerequisite for studying interactions at the level of perturbation theory as well as the back-reaction induced by a regularized stress-energy tensor.

The content of the paper is as follows. First, in Section \ref{sec: Horava-Lifshitz solutions}, we review the geometric data of the class of spacetimes of Lifshitz type that we considered. Secondly, in Section \ref{sec:
Klein-Gordon Field}, we consider the Klein-Gordon equation and we show that it can be reduced to a radial equation by a mode-decomposition of its solutions. Then, we study the radial equation together with its associated Green function, separately on Lifshitz spacetime, on a hyperbolic and on a spherical Lifshitz black hole, respectively in Sections \ref{sec: the
radial equation equation on Lifshitz spacetime, k=0}, \ref{sec: The radial equation on the hyperbolic Lifshitz black hole, k-1}
and \ref{sec: The radial equation on the spherical Lifshitz black hole, k+1}. In Section \ref{sec: Two-point functions for general boundary conditions}, we construct on each spacetime considered and for each admissible boundary condition the two-point function both of the ground and of a KMS/thermal state. Several computations of this work are reproduced in a Mathematica notebook available at \cite{github_notebook}.

\section{Ho\u{r}ava-Lifshitz solutions}
\label{sec: Horava-Lifshitz solutions}

In the following we call $\mathcal{R}$ the Ricci scalar determined  by a metric tensor $g$ on a four dimensional Lorentzian spacetime $\mathcal{M}$ while $L$ is scale length related to the cosmological constant via $\Lambda=-\frac{5}{L^2}$. In addition we indicate with $\varepsilon^{\mu\nu\alpha\beta}$ the Levi-Civita tensor density, with $F_{\mu\nu}=\partial_{[\mu}A_\nu]$ and
$H_{\mu\nu\tau}=\partial_{[\mu}B_\nu]$ two Abelian gauge fields, while $C=\frac{2}{L}$ is a constant, coupling parameter. As shown in \cite{Mann:2009yx}, four-dimensional Lifshitz topological black holes $(\mathsf{Lif}_\kappa,g_\kappa)$, where $\kappa\in\{-1,0,1\}$ are solutions of the Euler-Lagrange equations associated to the action
\begin{align}
  \label{eq: action Lifshitz gravity}
  S =\int_\mathcal{M} &d^4x\sqrt{|g|}\Big( \mathcal{R}-2\Lambda -\frac{1}{4}F_{\mu\nu}F^{\mu\nu}-\frac{1}{12}H_{\mu\nu\tau}H^{\mu\nu\tau} -\frac{C}{\sqrt{|g|}}\varepsilon^{\mu\nu\alpha\beta}B_{\mu\nu}F_{\alpha\beta} \Big),
\end{align}
where, given a global coordinate chart $(t,r,\theta,\varphi)$,
 \begin{equation}
   \label{eq: non-vanishing field strengths components}
       F_{rt} =  2 L r \text{  and   } H_{r\theta\varphi}= 2L^2 r j_\kappa(\theta),
 \end{equation}
with
\begin{subequations}
	\label{eq: metric Lifshitz black holes}
\begin{align}
	\label{eq: j_k}
	&j_\kappa(\theta) := \begin{cases}
		\theta,       & \kappa=0,\\
		\sinh(\theta),& \kappa=-1,\\
		\sin(\theta), & \kappa=+1.
	\end{cases}
\end{align}
\end{subequations}
All other components of $F$ or $H$ are either determined by these via antisymmetrization or they are vanishing. Here $t$ runs over the whole real line, $(\theta,\varphi)$ are the standard coordinates over the $2$-plane, the $2$-hyperboloid, and the $2$-sphere, respectively for $\kappa=0$, $\kappa=-1$ and $\kappa=1$. Also, $r\in(0,\infty)$ if $\kappa=0$ or $\kappa=1$, and $r\in(L/\sqrt{2},\infty)$ if $\kappa=-1$. For $\kappa\in\{-1,0,+1\}$, their line-element reads:

\begin{align}
  \label{eq: metric Lifshitz black holes no j}
    ds^2 =-\frac{r^{2}}{L^2} \left(\frac{r^{2}}{L^2} + \frac{\kappa}{2}\right) dt^2 + \left(\frac{r^{2}}{L^2} + \frac{\kappa}{2}\right)^{-1}dr^2
+  r^2 d\theta^2 + r^2j_\kappa(\theta)^2d\varphi^2,
\end{align}

This class of spacetimes share the scaling relations
\eqref{eq:scaling lifshitz} characteristic of Ho\u{r}ava-Lifshitz gravity
for $\underline{x}\in\{\theta,\varphi\}$ with $z=2$:
\begin{align}
  t\rightarrow \delta^2 t \text{, \quad}  r\rightarrow \frac{1}{\delta} r
 \text{\quad and \quad}   \underline{x}\rightarrow \delta \underline{x}.
\end{align}
All of them are static, non-globally hyperbolic and geodesically incomplete. For $\kappa=0$, $\kappa=-1$ and $\kappa=+1$, the line-element \eqref{eq: metric Lifshitz black holes} corresponds to, respectively, a flat, a hyperbolic and a spherical Lifshitz topological black hole. The Lifshitz horizon $r=0$ is a coordinate singularity for $\kappa=0$, but for $\kappa\neq 0$, the Ricci and the Kretschmann scalars diverge there. In the hyperbolic case, there is an horizon at $r=L/\sqrt{2}$, while in the spherical case, $r=0$ is a naked singularity. Comparing them with topological black holes within Einstein gravity \cite{Mann:1997iz}, we observe that compact sections at fixed $r$ and $t$ can also be obtained for Lifshitz topological black holes by suitable identifications in the spatial coordinates. In contrast, a naked singularity for the spherical black hole is present only in the Lifshitz case.

Furthermore, note that a flat topological black hole $(\mathsf{Lif}_0,g_0)$ is equivalent to a Lifshitz spacetime with critical exponent $z=2$ and with polar coordinates for the sections of constant time and radius. Hence, it also solves Einstein-Maxwell-Dilaton and Einstein-Proca gravity theories, as detailed in \cite[Pg.27]{Hartnoll:2016apf}.
\begin{figure}[H]
  \centering
   \includegraphics[align=c,width=.3\textwidth]{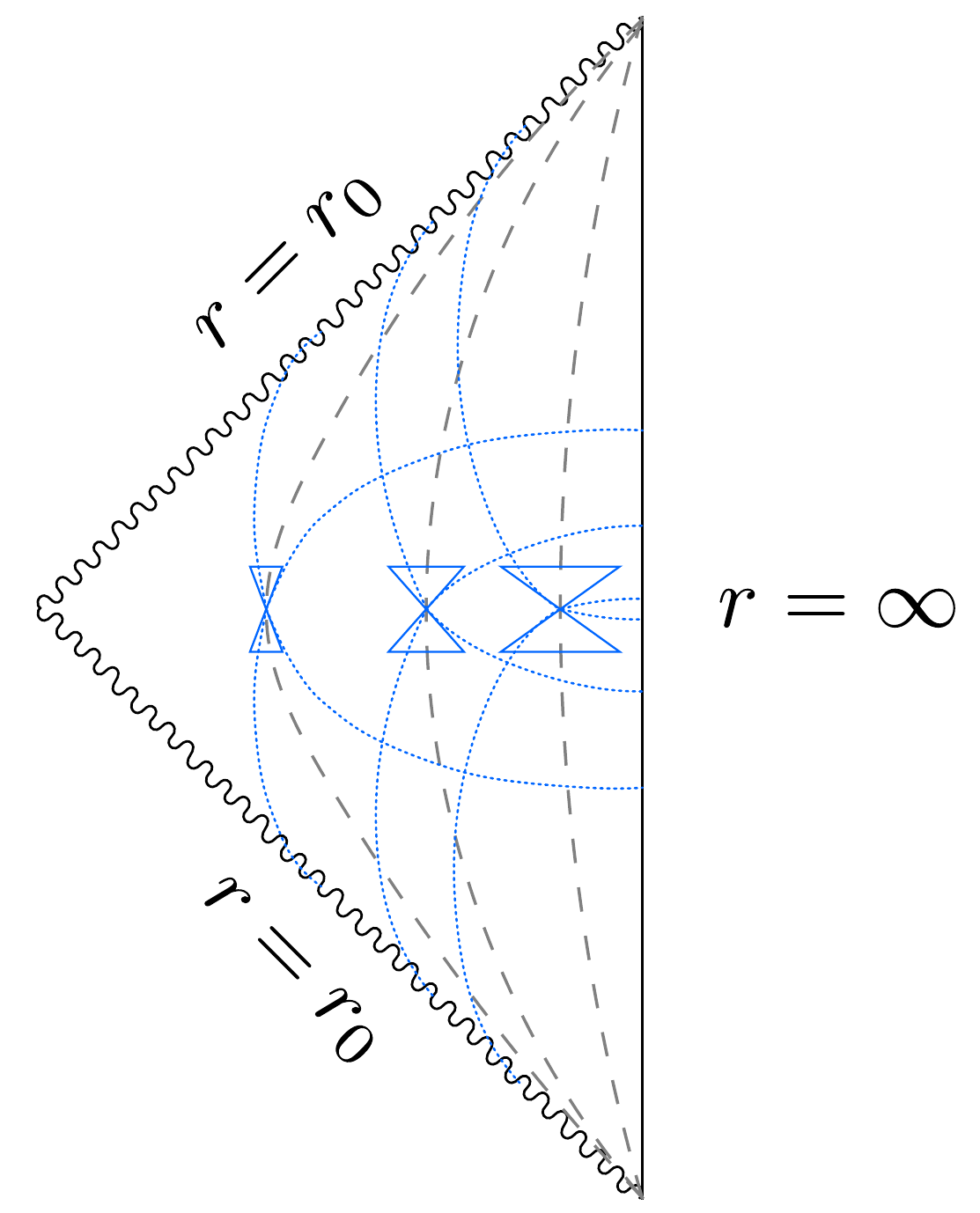}\hspace{.5cm}%
  \caption{Diagram representing flat ($r_0=0$), hyperbolic ($r_0=L/\sqrt{2}$) and spherical ($r_0=0$) Lifshitz topological black holes. In all cases we have a singularity at $r=r_0$ and a timelike boundary at $r=\infty$. The dashed lines are static trajectories, and the dotted lines are light rays.}
  \label{fig:diagram}
    \end{figure}
As $r\rightarrow \infty$, the $g_{tt}$ component of the line-element \eqref{eq: metric Lifshitz black holes} diverges faster than the others. This implies that near radial infinity the light cones flatten out and the effective speed of light diverges. Consequently, two distinct points at radial infinity with equal time share the same past and the same future, and, accordingly, we say that these spacetimes are not causally distinguishable at the boundary. An illustration of the causal structure--not a Penrose diagram in the standard sense---of Lifshitz topological black holes is given in Figure \ref{fig:diagram}. A detailed study regarding the anisotropic conformal infinity, generalizing Penrose's notion of conformal infinity, can be found in \cite{Horava:2009vy}. For elaborations on the metric-solutions \eqref{eq: metric Lifshitz black holes}, we refer to \cite{Mann:2009yx}.

\section{Klein-Gordon Field}
\label{sec: Klein-Gordon Field}

On $(\mathsf{Lif}_\kappa, g_\kappa)$, let us consider a real, massive, scalar field $\Psi : \mathsf{Lif}_\kappa \to \mathbb{R}$ whose action is given by
\begin{align}
  \label{eq: action scalar field}
S = -\frac{1}{2}\int_{\mathsf{Lif}_\kappa} d^4x\sqrt{|g_\kappa|} \, (\nabla_{\alpha} \Psi \nabla^{\alpha} \Psi + (\mu_0^2 + \xi\mathcal{R})\Psi^2),
\end{align}
where $\mu_0^2 \in \mathbb{R}^+$ is the mass parameter of the scalar field, $\xi \in \mathbb{R}$ is the scalar-curvature coupling constant and the
Ricci scalar built out $g_\kappa$ is given by:
\begin{equation}
  \mathcal{R}= -\frac{22}{L^2} - \frac{\kappa}{r^2},
\end{equation}
where, with a slight abuse of notation, we omit to indicate explicitly the $\kappa$-dependence of $\mathcal{R}$. For convenience, we define the effective mass $\mu^2 = \mu_0^2 - \xi \frac{22}{L^2}  \in \mathbb{R}$. Observe that, due to the anisotropy of the line-element, we cannot rescale the coordinates to obtain a constant Ricci scalar.

From the action \eqref{eq: action scalar field}, it descends the Klein-Gordon equation, which governs the dynamics of the scalar field:
\begin{equation}
\label{eq: KG equation general}
  P\Psi = (\Box - \mu^2 +  \xi \frac{\kappa}{r^2}) \Psi = 0,
\end{equation}
where $\Box$ is the D'Alembert wave operator built out of the metric $g_\kappa$ on $\mathsf{Lif}_\kappa$.

In order to solve Equation \eqref{eq: KG equation general}, with line-element given by Equation \eqref{eq: metric Lifshitz black holes}, we observe that the metric admits a global timelike Killing field $\partial_t$ and
the scalar field $\Psi$ can be written in terms of a Fourier expansion with respect to $t$. Moreover, by separation of variables and the superposition principle, we make the following ansatz:
\begin{align}
\label{eq: Fourier expansion of scalar field}
\Psi(t,r,\theta,\varphi) = \int\limits_{\sigma(\triangle)}d\Sigma(\ell,m)\int\limits_{\mathbb{R}}d \omega\, e^{-i\omega t } R(r)Y_\ell^m(\theta,\varphi).
\end{align}
The harmonics $Y_\ell^m(\theta,\varphi)$ are the eigenfunctions of the Laplacian operator $\triangle$ with eigenvalue $\lambda_\ell^m$. The Laplacian operator comes from the angular part of $\Box$ and it depends on the case considered:
	\begin{equation}
	\triangle = \begin{cases}
	         \frac{1}{\theta^2}\frac{\partial^2}{\partial \varphi^2} + \frac{1}{\theta}\frac{\partial}{\partial\theta}+\frac{\partial^2}{\partial \theta^2}, &\text{ for }\kappa=0; \\
		 \frac{1}{\sinh(\theta)^2}\frac{\partial^2}{\partial \varphi^2} + \frac{1}{\sinh(\theta)}\frac{\partial}{\partial\theta}\left(\sinh(\theta )\frac{\partial}{\partial\theta}\right), &\text{ for }\kappa=-1; \\
	\frac{1}{\sin(\theta)^2}\frac{\partial^2}{\partial \varphi^2} + \frac{1}{\sin(\theta)}\frac{\partial}{\partial\theta}\left(\sin(\theta )\frac{\partial}{\partial\theta}\right), &\text{ for }\kappa=+1.
	\end{cases}
	\end{equation}
The integral with measure $d\Sigma(\ell,m)$ over the space $\sigma(\triangle)$ denotes that we are summing over all harmonics, which depend on the
quantum numbers $\ell$ and $m$. For the explicit definition of the harmonics, of $d\Sigma(\ell,m)$ and of $\sigma(\triangle)$, we refer the reader
to \cite{terras2012harmonic} for the flat and spherical cases, and to \cite{Limic} or to \cite[Appx.A]{Campos:2020lpt} for the hyperbolic case. On
the contrary, since they appear explicitly in the form of the radial component $R(r)$, we report explicitly the form of the eigenvalues $\lambda^m_\ell$:
  \begin{align}
  \label{eq: definition lambda}
        \lambda_\ell^m = \begin{cases}
                        -(\ell^2 + m^2),\, \ell,m\in\mathbb{R}, &\text{ for }\kappa=0,\\
                        -(\frac{1}{4}+\ell^2),\, \ell\in\mathbb{R}, &\text{ for }\kappa=-1,\\
                        -\ell(\ell+1),\, \ell\in\mathbb{N}_0, &\text{ for
}\kappa=+1.
                    \end{cases}
  \end{align}
 It follows that $\lambda_\ell^m$ ranges over a discrete spectrum for $\kappa=+1$, while, otherwise, it assumes continuous values on the negative portion of the real line.

Substituting the ansatz \eqref{eq: Fourier expansion of scalar field} in Equation \eqref{eq: KG equation general}, the Klein-Gordon equation reduces to an ordinary differential equation, which we call \textit{the radial
equation}
\begin{subequations}
  \label{eq: the radial equation}
\begin{equation}
    R''(r)+ Q(r)R'(r)+ V(r)R(r)=0,
\end{equation}
with
\begin{align}
\label{eq: the radial equation coefficient functions}
     Q(r)= &\frac{4 r}{\kappa L^2+2 r^2}+\frac{3}{r},\\
     V(r)= &\frac{4 L^6}{\left(\kappa L^2 r+2 r^3\right)^2} \omega ^2 +
\frac{2 L^2}{\kappa L^2 r^2+2 r^4}\lambda -\frac{2 L^2}{\kappa L^2+2 r^2}
 \mu^2,
\end{align}
where
\begin{equation}
  \label{eq: lambda shited by xi}
  \lambda := \lambda_\ell^m + \kappa \xi.
\end{equation}
\end{subequations}
It is noteworthy that the radial potential associated to the Ricci scalar
leads solely to a shift of the eigenvalue $\lambda_\ell^m$ by a factor $\kappa\xi$. For $\kappa=0$ and $\kappa=+1$, the domain of the radial equation is $r\in(0,\infty)$, while for $k=-1$, we have $r\in(L/\sqrt{2},\infty)$. According to the standard endpoint classification of ordinary differential equations, in all cases, $r=\infty$ is an irregular singular point. The other endpoint is a regular singularity for $\kappa=\pm1$, while it is an irregular one for $\kappa=0$.

The radial equation can be written in well-known forms: as a hypergeometric equation for $\kappa\neq 0$ and as a confluent hypergeometric equation
for $\kappa=0$. To obtain these forms we apply different coordinate changes in each case, hence we treat them separately in the following sections. Before that, some general considerations are in order.

First of all, observe that, since the geometric structure of radial infinity is the same on all $\mathsf{Lif}_\kappa$ spacetime, the solutions of Equation \eqref{eq: the radial equation} behave similarly asymptotically,
namely
\begin{subequations}
\begin{align}
  \label{eq: asymptotic R(r) at radial infinity }
  &R(r)\sim r^{-2 \pm \nu}   \text{, as  }r\rightarrow \infty,
\end{align}
where
\begin{align}
  \label{eq: definition of nu}
  &\nu := \sqrt{4 + L^2 \mu^2}.
\end{align}
\end{subequations}
We require $\nu >0$, which is equivalent to imposing the Breitenlohner-Freedman bound on the effective mass. In addition, observe that negative values of the effective mass encompass
\begin{equation}
  \label{eq: coupling for negative effective mass}
  \mu^2\in \left( -\frac{4}{L^2}, 0 \right) \iff \xi \in \left( \frac{L^2\mu_0^2}{22}, \frac{L^2\mu_0^2 +4}{22} \right)\iff \nu\in(0,2).
\end{equation}
 In particular, the massless conformally-coupled case corresponds to $\mu_0^2=0$ and $\xi=\frac{1}{6}\in\left(0, \frac{4}{22} \right)$, which entails $\mu^2 = - \frac{22}{6 L^2}$ and, accordingly, $\nu=\frac{1}{\sqrt{3}}$.

A direct inspection of Equation \eqref{eq: asymptotic R(r) at radial infinity } unveils the existence of two different asymptotic behaviors at infinity. Hence, in order to select one among the infinite possible solutions at $r\to \infty$, one needs to identify a suitable criterion. This is the core of the next sections and it relies on the observation that Equation \eqref{eq: the radial equation} can be rewritten as an eigenvalue problem for a Sturm-Liouville operator, {\em i.e.}
\begin{subequations}
\begin{equation}
  \label{eq: Sturm-Liouville form R(r)}
        L_{\omega^2}  R(r) = \omega ^2 R(r),
\end{equation}
where
\begin{equation}
  \label{eq: Sturm-Liouville operator L}
  L_{\omega^2}:=-\frac{1}{q(r)} \left(\frac{d}{d r} \left(p(r)\frac{d}{d r}\right)+v(r) \right),
\end{equation}
with coefficient functions:
    \label{eq: Sturm-Liouville coefficient functions r}
\begin{align}
  &q(r) = \frac{4 L^6 r}{\kappa L^2+2 r^2}, \label{eq:measure mu(r)}\\
  &p(r) = \kappa L^2 r^3+2 r^5, \label{eq:SL operator function p(r)}\\
  &v(r) = 2 L^2 r \left(-\mu^2 r^2+\lambda \right).
\end{align}
\end{subequations}

\noindent For later convenience, observe that the Sturm-Liouville operator satisfy the following properties under conjugation and reflection of $\omega\in\mathbb{C}$:
\begin{subequations}
\begin{align}
  \label{eq: properties of Sturm-Liouville operator L_omega}
    & L_{\overline{\omega}^2} =  \overline{L_{\omega^2}},\\
    & L_{(-\omega)^2} = L_{(\omega)^2}.
\end{align}
\end{subequations}

\section{The radial equation on Lifshitz spacetime}
\label{sec: the radial equation equation on Lifshitz spacetime, k=0}
In this section, we study the solutions of the radial equation \eqref{eq:
the radial equation} on $\mathsf{Lif}_0$ using the tools proper of Sturm-Liouville theory, see {\em e.g.} \cite{Zettl:2005}. First, we show that these solutions can be constructed starting from an associated confluent hypergeometric equation. Subsequently, in Section \ref{sec: The radial solutions k=0} we choose a convenient basis of solutions and, in Section \ref{sec: Square-integrability conditions k=0}, we analyze whether they lie in a suitably specified Hilbert space. This analysis will highlight whether at radial infinity we can impose generalized Robin boundary conditions, as detailed in Section \ref{sec: Robin boundary conditions at r=infty k 0}.
As a by product, in Section \ref{sec: The radial Green function k 0} we construct the radial Green function. Finally, in Section \ref{sec: on the existence of bound states k=0}, we study the existence of bound states,
which is tantamount to determining which among the admissible boundary conditions yield physically sensible two-point functions. We emphasize that
a construction and an analysis similar to the following one has been already applied for a scalar field on an anti-de Sitter spacetime in \cite{Dappiaggi:2016fwc,Dappiaggi:2018xvw}, on a rotating BTZ black hole spacetime in \cite{Dappiaggi:2018pju} and on a massless topological black hole in
\cite{Campos:2020lpt}. In particular, here we follow the same nomenclature and strategy outlined in \cite{Dappiaggi:2016fwc} to which we refer for
further details.

\subsection{The radial equation as a confluent hypergeometric equation}
\label{sec: The radial equation as a confluent hypergeometric equation}
\noindent If one considers the coordinate change
\begin{equation}
  r\mapsto u = \frac{iL^3\omega}{r^2}\in(0,\infty),
\end{equation}
the radial equation \eqref{eq: the radial equation} for $\kappa=0$ reads
\begin{equation}
  \label{eq: radial equation R(u) k=0}
      u R''(u) -R'(u)  - \frac{1}{4} \left(\frac{L^2 \mu^2}{u}+ \frac{i \lambda }{L \omega }+ u\right) R(u)=0,
\end{equation}
At radial infinity, which is now located at $u=0$, $R(u)\sim u^{1\pm\frac{\nu}{2}}$ in agreement with Equation \eqref{eq: asymptotic R(r) at radial infinity }. At $u=\infty$, we find that $R(u)\sim e^{\pm u/2}$. Thus, we make the ansatz
\begin{align}
  \label{eq: ansatz R(u) k=0}
    &R(u) =  e^{-u/2} \left(\frac{u}{i L^3 \omega}\right)^{\frac{1}{2} \left(2 + \nu \right)} w(u).
\end{align}
It follows that this is a solution of Equation \eqref{eq: radial equation
R(u) k=0} if and only if $w(u)$ is a solution of the confluent hypergeometric equation
\begin{subequations}
  \label{eq: confluent hypergeometric w(u) k=0}
\begin{align}
  u w''(u) + (b_0 - u)w'(u) -a_0 w(u) =0,
\end{align}
with
\begin{align}
  \label{eq: parameters of confluent hypergeometric w(u) k=0}
    &a_0 = \frac{1+\nu}{2} + \frac{i \lambda }{4 L \omega },\\
    &b_0 = 1+ \nu .
\end{align}
\end{subequations}

\subsection{The radial solutions}
\label{sec: The radial solutions k=0}
In view of the results of the previous section, a basis $\{R_{1(u_0)}(u),
R_{2(u_0)}(u)\}$ of solutions of the radial equation near a point $u_0$ can be written in terms of a basis $\{w_{1(u_0)}(u), w_{2(u_0)}(u)\}$ of Equation \eqref{eq: confluent hypergeometric w(u) k=0}. That is, for $j\in\{1,2\}$ we define
\begin{align}
  \label{eq: basis R(u) k=0}
    &R_{j(u_0)}(u) =  e^{-u/2} \left(\frac{u}{i L^3 \omega}\right)^{\frac{1}{2} \left(2 + \nu \right)} w_{j(u_0)}(u).
\end{align}
In particular we are interested in the case $u_0=0$ or $u_0\to\infty$. When $b_0 \notin \mathbb{Z}$, satisfactory solutions of \eqref{eq: confluent hypergeometric w(u) k=0} near the endpoint $u_0$ are given in Table
\ref{tab: basis confluent hypergeo}

 \begin{table}[H]
   \centering
   \caption{Suitable bases of the confluent hypergeometric equation \eqref{eq: confluent hypergeometric w(u) k=0}.}\smallskip
   \label{tab: basis confluent hypergeo}
   \includegraphics[width=.7\textwidth]{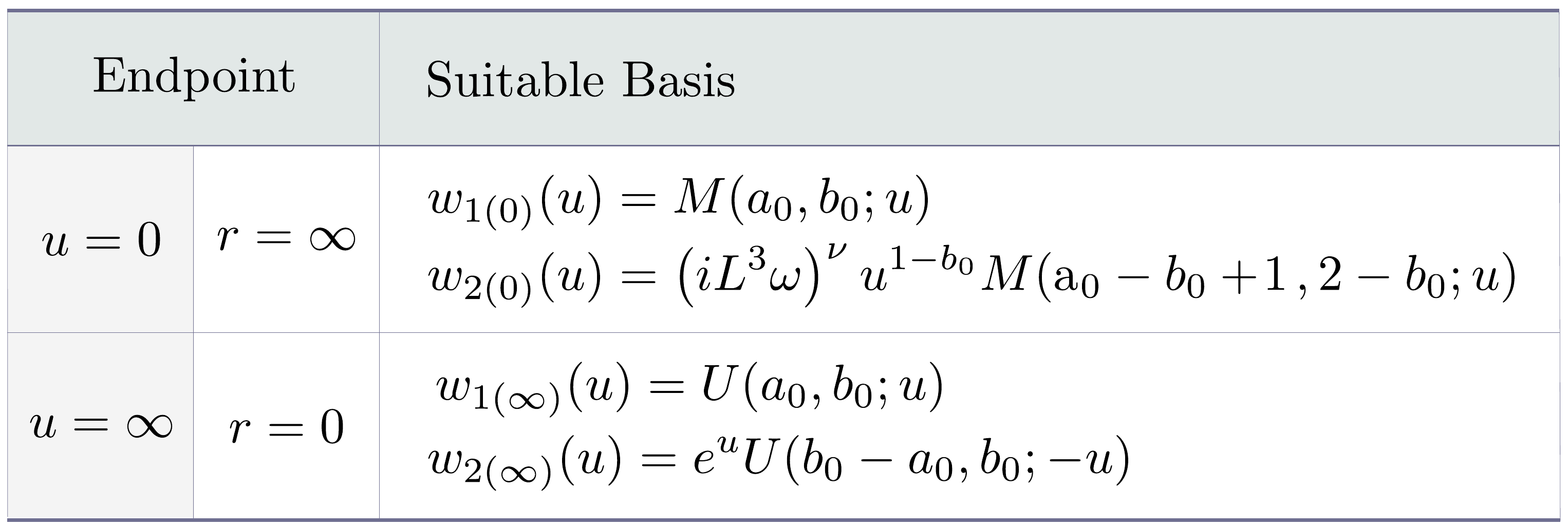}
  \end{table}

The confluent hypergeometric functions $M$ and $U$, which are also known as Kummer's functions, are discussed in \cite[Ch.13]{NIST}. As detailed in Section 13.2 of \cite[Ch.13]{NIST}, when the parameter $b_0$ assumes integer values, and depending on the value of the parameter $a_0$, we have to replace the solution $w_{2(0)}$ as stated in Table \ref{tab: basis confluent hypergeo}. However, whenever this is the case, the new solution does not abide to a necessary square-integrabilty condition which we will discuss in the next section. For this reason we omit giving their explicit
expressions.

\subsection{Square-integrability conditions}
\label{sec: Square-integrability conditions k=0}

 Following the rationale of Sturm-Liouville theory -- see in particular \cite{Dappiaggi:2016fwc, Zettl:2005} --, we are looking for solutions $R(r)$ of the radial equation such that there exists a neighbourhood of
$r_0$ for which $R(r)$ is square integrable with respect to the measure $q(r)$ as defined in Equation \eqref{eq:measure mu(r)}.
Next in order, we verify such property for the radial solutions \eqref{eq: basis R(u) k=0}, written in terms of the solutions of Table \ref{tab: basis confluent hypergeo}, hence classifying the endpoints of the domain of the radial equation with $\kappa=0$ into limit point or limit circle, according to Weyl's endpoint classification, see \cite{Zettl:2005}.

  At $u=0$ ($r=\infty$), the solution $w_{1(0)}(u)$ reduces to $1$, while $w_{2(0)}(u)$ tends asymptotically to $u^{1-b_0}$. Therefore, the radial solutions \eqref{eq: basis R(u) k=0}, written in terms of $r$ instead of $u$, behave asymptotically as:
  \begin{subequations}
  \begin{align}
    R_{1(\infty)}(r) \sim r^{-2-\nu},\\
    R_{2(\infty)}(r) \sim  r^{-2+\nu}.
  \end{align}
\end{subequations}
  Taking into account the measure \eqref{eq:measure mu(r)}, one finds that both solutions lay in $L^2((r_0,\infty);q(r))$, $\forall r_0 >0$, if and only if $\nu \in (0,1)\cup(1,2)$. For this range of values of $\nu$, $r=\infty$ is a limit circle. For $\nu=1$ or $\nu \geq 2$, only $ R_{1(0)}(r)$ is square-integrable and $r=\infty$ is a limit point.

On the other hand, the endpoint $r=0$ is always a limit point. By direct inspection, it follows that
\begin{subequations}
\begin{align}
|R_{1(0)}(r)|^2 \sim  e^{+\frac{L^3}{r^2}\Imag(\omega )},\\
|R_{2(0)}(r)|^2 \sim  e^{-\frac{L^3}{r^2}\Imag(\omega )}.
\end{align}
\end{subequations}
Therefore, the only solution lying in $L^2((0,r_0);q(r))$ for all $r_0<\infty$ is
\begin{align}
  \label{eq: R_0 for k=0}
 R_{0}(r) = \begin{cases}
                      R_{1(0)}(r) , \text{ for }\Imag(\omega) < 0, \\
                      R_{2(0)}(r) , \text{ for }\Imag(\omega) > 0.
                    \end{cases}
\end{align}

\subsection{Symmetries under conjugation and reflection}
\label{sec: Symmetries under conjugation and reflection k=0}

In this section, we show that the properties of the Sturm-Liouville operator as per Equation \eqref{eq: properties of Sturm-Liouville operator L_omega} reflect themselves in those of the solutions \eqref{eq: basis R(u) k=0} of the radial equation. First, let us make the $\omega$-dependence
explicit, by denoting the parameter $a_0$ by $a_0(\omega)$ and defining the auxiliary function:
\begin{equation}
 \alpha(r,\omega)=e^{- \frac{iL^3\omega}{2r^2}} r^{-2 - \nu}.
\end{equation}
The suitable basis at the endpoint $r=\infty$ reads:
\begin{subequations}
  \label{eq: basis at u=0 R(u) k=0 Explicitly}
\begin{align}
&R_{1(\infty)}(r,\omega) =   \alpha(r,\omega) M\left(a_0(\omega),b_0;iL^3\omega r^{-2}\right),\\
&R_{2(\infty)}(r,\omega) =   \alpha(r,\omega)r^{2\nu} M\left(a_0(\omega)-b_0+1,2-b_0; iL^3\omega r^{-2}\right),
\end{align}
\end{subequations}
\noindent and at $r= 0$:
\begin{subequations}
  \label{eq: basis at u=infty R(u) k=0 Explicitly}
\begin{align}
&R_{1(0)}(r,\omega) = \alpha(r,\omega) U\left(a_0(\omega),b_0;iL^3\omega r^{-2}\right),\\
&R_{2(0)}(r,\omega) =  \alpha(r,\omega)e^{\frac{iL^3\omega}{r^2}} U\left(b_0-a_0(\omega),b_0;-iL^3\omega r^{-2}\right).
\end{align}
\end{subequations}
From the definition of the parameters $a_0(\omega)$ and $b_0$, see Equation \eqref{eq: parameters of confluent hypergeometric w(u) k=0}, it follows
\begin{subequations}
  \label{eq: properties of the parameters of confluent hypergeometric w(u) k=0}
\begin{align}
  & a_0(\overline{\omega}) =  \overline{b_0 - a_0(\omega)}, \\
  &  a_0(-\omega) =  b_0 - a_0(\omega).
\end{align}
\end{subequations}
Taking these properties into account \eqref{eq: properties of the parameters of confluent hypergeometric w(u) k=0} together with the Kummer's transformation:
\begin{align}
\label{eq: Kummer transformation M}
  & M(a_0(\omega),b_0,u) = e^{u}M(b_0-a_0(\omega),b_0,-u),
\end{align}
we obtain, for $j_1\in\{1,2\}$, that the basis at $r=\infty$ satisfies
\begin{align}
\label{eq: properties basis R(r) at infty k=0}
&R_{j_1(\infty)}(r,\overline{\omega}) = \overline{R_{j_1(\infty)}(r,\omega)}.
\end{align}
Also, for $j_2\in\{1,2\}$, with $j_1\neq j_2$, we find that the basis at $r=0$ satisfy:
\begin{align}
  \label{eq: properties basis R(r) at 0 k=0}
&R_{j_1(0)}(r,\overline{\omega}) = \overline{R_{j_2(0)}(r,\omega)}.
\end{align}
Furthermore, since the solutions depend on $\omega$ only through a factor
$i\omega$, the conjugation properties imply that for frequencies such that $\omega^2\in\mathbb{R}$, the solutions at $r=\infty$ are real-valued.
At the same time, the solutions at $r=0$ are not necessarily real-valued for real frequencies, but in this case one is the complex conjugate of the other.

\subsection{Robin boundary conditions at $r=\infty$}
\label{sec: Robin boundary conditions at r=infty k 0}

In the range $\nu \in (0,1)\cup(1,2)$, both solutions are square-integrable at radial infinity. Hence, in the spirit of Sturm-Liouville theory, in
this case an admissible solution is a linear combination of $R_{1(\infty)}(r)$ and $R_{2(\infty)}(r)$. In fact, there exists a one-parameter family of suitable boundary conditions such that, for each $\gamma\in[0,\pi)$,
the solution
  \begin{align}
    \label{eq: R_gamma}
  R_{\gamma}(r)= \cos(\gamma)R_{1(\infty)}(r) + \sin(\gamma)R_{2(\infty)}(r)
  \end{align}
satisfies
  \begin{align}\label{1.29}
  \lim_{r \to \infty}\left(\cos(\gamma)W_r[R_{\gamma}(r), R_{1(\infty)}(r)] + \sin(\gamma)W_r[R_{\gamma}(r), R_{2(\infty)}(r)]\right) = 0,
  \end{align}
where we recall that, given two differentiable functions $ f_1(x)$ and $f_2(x)$ defined on an interval $I\subseteq\mathbb{R}$, the Wronskian $W$ reads
  \begin{align}
  W_x\left[f_1(x),f_2(x)\right] = f_1(x)\frac{df_2(x)}{dx} - \frac{df_1(x)}{dx}f_2(x).
  \end{align}
For $\gamma = 0$ we recover Dirichlet boundary condition, given that $R_{\gamma}(r)$ reduces to the principal solution $R_{1(\infty)}$. For $\gamma = \frac{\pi}{2}$ we have Neumann boundary condition, for a fixed secondary solution $R_{2(\infty)}$. The cases $\gamma \in \left(0,\frac{\pi}{2}\right)\cup \left(\frac{\pi}{2},\pi\right)$, are referred to as Robin
boundary conditions. In reality, Dirichlet and Neumann can be seen as particular cases of Robin boundary conditions, but here we make the distinction for convenience in future referencing.

Given that the basis chosen at $r=\infty$ satisfy properties \eqref{eq:
properties basis R(r) at infty k=0}, it holds:
\begin{align}
 \label{eq:symmetry conjugation Rgamma}
R_{\gamma}(r,\overline{\omega})=\overline{R_{\gamma}(r,\omega)}.
\end{align}

\subsection{The radial Green function}
\label{sec: The radial Green function k 0}

Using the results of the previous sections we can construct the Green's distribution $\mathcal{G}_{\omega}(r,r')$ of the differential operator $L_{\omega^2}$, given by Equation \eqref{eq: Sturm-Liouville operator L}, {\em i.e.} a solution of
  \begin{equation}
    \label{1gpr990b0}
    ((L_{\omega^2}-\omega^2) \otimes \mathbbm{1})\mathcal{G}_\omega(r,r')
= (\mathbbm{1} \otimes (L_{\omega^2}-\omega^2))\mathcal{G}_\omega(r,r')
= \frac{\delta(r-r')}{q(r)}.
  \end{equation}
Consider the range of values $\nu \in (0, 1)\cup(1,2)$. Following \cite{greenBook}, we are able to write $\mathcal{G}_{\omega}(r,r')$ in terms of solutions of the radial solutions $R_0$ and $R_{\gamma}$, as in \eqref{eq: R_0 for k=0} and \eqref{eq: R_gamma}, as
  \begin{subequations}
    \label{eq: green function k 0}
  \begin{align}
  \mathcal{G}_{\omega}(r,r') =\frac{1}{ \mathcal{N}_{\omega}}\left( \Theta(r'-r) R_{0}(r)R_{\gamma}(r') + \Theta(r-r')R_{0}(r')R_{\gamma}(r)\right).
  \end{align}
For $p(r)$ as in Equation \eqref{eq:SL operator function p(r)}, the normalization $ \mathcal{N}_{\omega}$ is
  \begin{equation}
   \label{eq: definition normalization of radial green function k=0}
  \mathcal{N}_{\omega} := -p(r)W_r\left[ R_{0}(r),R_{\gamma}(r)\right].
  \end{equation}
  \end{subequations}
The solution $R_0$ is given by Equation \eqref{eq: R_0 for k=0}, while $R_\gamma$ is given by Equation \eqref{eq: R_gamma} together with Eqution
\eqref{eq: basis at u=0 R(u) k=0 Explicitly}. Assuming $\Imag(\omega)<0$, we have $R_0(r)=R_{1(0)}(r)$. Considering the fundamental relation
that connects the solution $R_{1(0)}(r)$ with the solutions at radial infinity
  \begin{subequations}
  \begin{equation}
  \label{eq: fundamental relation R10 as R1infty R2infty k=0}
  R_{1(0)}(r) = A_0 R_{1(\infty)}(r) + B_0 R_{2(\infty)}(r),
  \end{equation}
with coefficients
  \begin{align}
    &A_0:= \frac{\Gamma(1-b_0)}{\Gamma(a_0-b_0+1)},\\
    &B_0:= \frac{\Gamma(b_0-1)}{\Gamma(a_0)}(iL^3\omega)^{-\nu},
  \end{align}
  \end{subequations}
and taking into account the Wronskian \cite[(13.2.33)]{NIST}
  \begin{equation}
    W_u[w_{1(0)},w_{2(0)}] = (i L^3\omega)^\nu(1-b_0) u^{-b_0}e^u,
  \end{equation}
the normalization \eqref{eq: definition normalization of radial green function k=0} reads
  \begin{equation}
  \label{eq: normalization k = 0 omega A0 and B0}
  \mathcal{N}_\omega=4\nu\{B_0\cos(\gamma) - A_0\sin(\gamma)\}, \text{ for }\Imag{\omega}<0.
  \end{equation}
With respect to frequencies $\overline{\omega}$ with $\Imag(\overline{\omega})>0$, we take $R_0(r)=R_{2(0)}(r)$ and the normalization in the upper part of the complex plane, by an analogous computation and making the frequency dependence of $a_0$ explicit, reads
 \begin{equation}
 \label{eq: normalization k = 0 conjugate omega}
 \mathcal{N}_{\overline{\omega}} = 4\nu \left\{  \left(-i L^3\overline{\omega} \right)^{-\nu} \cos(\gamma)\frac{\Gamma(b_0-1)}{\Gamma(b_0-a_0(\overline{\omega}))} - \sin(\gamma)\frac{\Gamma(1-b_0)}{\Gamma(1-a_0(\overline{\omega}))}\right\}, \text{ for }\Imag{\overline{\omega}}>0.
 \end{equation}
The relations individuated in Equation \eqref{eq: properties of the parameters of confluent hypergeometric w(u) k=0} imply that, with respect to
complex conjugation, the normalization and the radial Green function satisfy
    \begin{align}
      \label{eq: property normalization and green omega conjugate k=0}
    \mathcal{N}_{\overline{\omega}} &= \overline{\mathcal{N}_{\omega}} \quad \text{ and }\quad\mathcal{G}_{\overline{\omega}}=\overline{\mathcal{G}_{\omega}}.
    \end{align}
Therefore, we can limit ourselves to study in detail either the case $\Imag(\omega)<0$ or $\Imag(\omega)>0$, and then, by symmetry, extend the results to the other case. Next, we compute the normalization and we verify the existence of bound states, assuming  $\Imag(\omega)<0$.

\subsection{On the existence of bound states}
\label{sec: on the existence of bound states k=0}

    Consider $\Imag(\omega) < 0$. The poles of the radial Green function are the zeros of the normalization constant. In the following, we state the conditions that yield $\mathcal{N}_{\omega}=0$ for Dirichlet, Neumann and Robin boundary conditions. We recall that we are considering $\nu\in(0,1)\cup(1,2)$, $n\in\mathbb{N}_0$ and that the Gamma function satisfies
 \begin{equation}
   \frac{1}{\Gamma(z)}=0\iff z\in\mathbb{Z}^-_0.
 \end{equation}
  \subsubsection{Dirichlet boundary condition, $\gamma=0$}
\vspace{-1cm}
 \begin{align}
  &\mathcal{N}_{\omega}=0 \iff a_0=-n  \iff \omega = -i\frac{\lambda}{2L(1+\nu+2n)}.
 \end{align}
The frequencies above have $\Imag(\omega)>0$, because $\lambda\leq0$, thus they do not lie in the domain of $\mathcal{N}_{\omega}$ and can be neglected.

\subsubsection{Neumann boundary condition, $\gamma=\pi/2$}
  \vspace{-1cm}
 \begin{align}
  & \mathcal{N}_{\omega}=0 \iff  a_0-b_0+1 = -n
   \iff \omega = -i\frac{\lambda}{2L(1-\nu+2n)}.
 \end{align}
 For $\nu\in(0,1)$, $1-\nu+2n>0$ and the frequencies above have positive imaginary part. At the same time, for $\nu\in(1,2)$ and $n=0$, the denominator is negative, and for each $\lambda$ there is a unique pole with negative imaginary part at
 \begin{equation}
    \omega_p = -i\frac{\lambda}{2L(1-\nu)}.
 \end{equation}
  \subsubsection{Robin boundary conditions, $\gamma\in(0,\pi/2)\cup(\pi/2,\pi)$}
 \noindent Define
 \begin{subequations}
 \begin{align}
    &\chi(\omega) := \tan(\gamma) - \Xi(\omega), \label{eq:chi = tan - Xi} \\
    &\Xi(\omega) := \left(i L^3\omega \right)^{-\nu}\frac{\Gamma(a_0-b_0+1)}{\Gamma(a_0)}\frac{\Gamma(b_0-1)}{\Gamma(1-b_0)} \label{eq: Xi simpler}
 \end{align}
\end{subequations}
 The zeros of the normalization are solutions of
 \begin{equation}
   \label{eq: transcendental eq k=0}
   \chi(\omega)=0.
 \end{equation}
The zeros and poles of $\Xi(\omega)$ are, respectively, the zeros of the normalization for Dirichlet and Neumann boundary conditions. Thus its zeros all lie in the upper half on the $\omega$-complex plane and it has no poles for $\nu\in(0,1)$, and one pole per $\lambda$ for $\nu\in(1,2)$.

\begin{figure}[H]
   \centering
    \includegraphics[align=c,width=.4\textwidth]{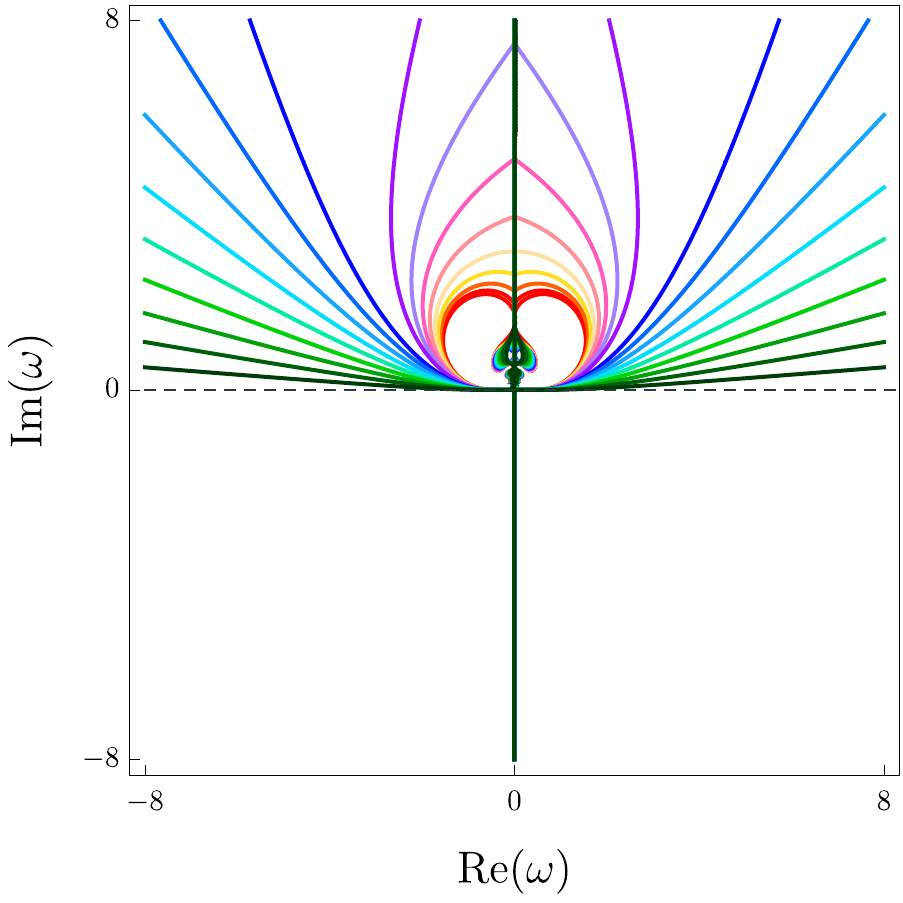}\hspace{.5cm}%
    \includegraphics[align=c,width=.1\textwidth]{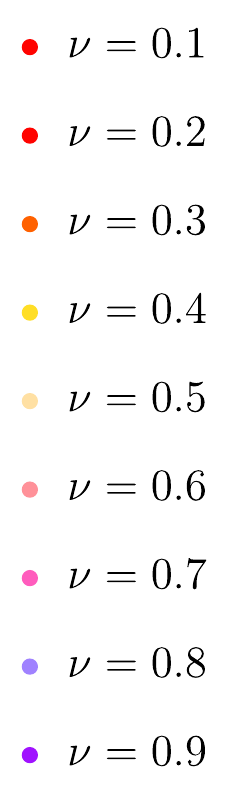}\hspace{.5cm}%
    \includegraphics[align=c,width=.1\textwidth]{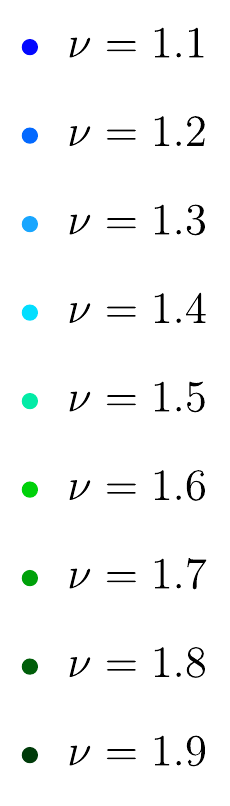}
   \caption{Contour lines of $\Imag(\Xi)=0$ for $L=1$, $\lambda=-3$
and several values of $\nu$. For $\Imag(\omega)<0$, only $\omega=i\Imag(\omega)$ is present.}
   \label{fig:Im(Xi) on the complex plane k = 0}
     \end{figure}

Although Equation \eqref{eq: transcendental eq k=0} is transcendental, it always admits a solution whenever $\Xi(\omega)$ is real-valued. In addition, the set of frequencies that solves Equation \eqref{eq: transcendental eq k=0} lies in the point spectrum of the differential operator determined by the radial part of the Klein-Gordon operator. Since we consider solely self-adjoint extensions, its $\omega^2$-spectrum is necessarily within $\mathbb{R}$. Therefore, we conclude that $\mathcal{N}_\omega=0\iff\omega=-i|\Imag(\omega)|$, as illustrated in Figure \ref{fig:Im(Xi) on the complex plane k = 0}. Observe that the argument above does not apply for $\Imag(\omega)>0$, since the upper part of the $\omega$-complex plane is not in the domain of $\mathcal{N}_\omega$, and in fact, we notice that it has more lines of real-phase for frequencies with positive imaginary part.
Next, we analyze the behavior of $\Xi$ along the negative imaginary axis,
i.e. for $\omega=i\Imag(\omega)$ with $\Imag(\omega)<0$, for $\nu\in(0,1)$ and $\nu\in(1,2)$, separately. Observe that
\vspace{-.5cm}
 \begin{subequations}
 \begin{align}
   &\lim\limits_{\omega\rightarrow 0} \Xi( i \omega) = \left(\frac{-\lambda L^2}{4}\right)^{-\nu}\frac{\Gamma(\nu)}{\Gamma(-\nu)}, \label{eq: limits of Xi k=0 at omega 0}\\
   &\lim\limits_{\omega\rightarrow \omega_p} \Xi(\omega) = \infty , \label{eq: limits of Xi k=0 at omega omegap}\\
   &\lim\limits_{\omega\rightarrow \infty} \Xi(- i \omega) = 0 ,\label{eq: limits of Xi k=0 at omega infty}
 \end{align}
 \end{subequations}
and note that the right hand side of Equation \eqref{eq: limits of Xi k=0 at omega 0} is negative if $\nu\in(0,1)$, and positive if $\nu\in(1,2)$. \\

 \noindent\underline{The analysis for $\nu\in(0,1)$:}\\

\begin{figure}[H]
 \centering
   \includegraphics[width=.9\textwidth]{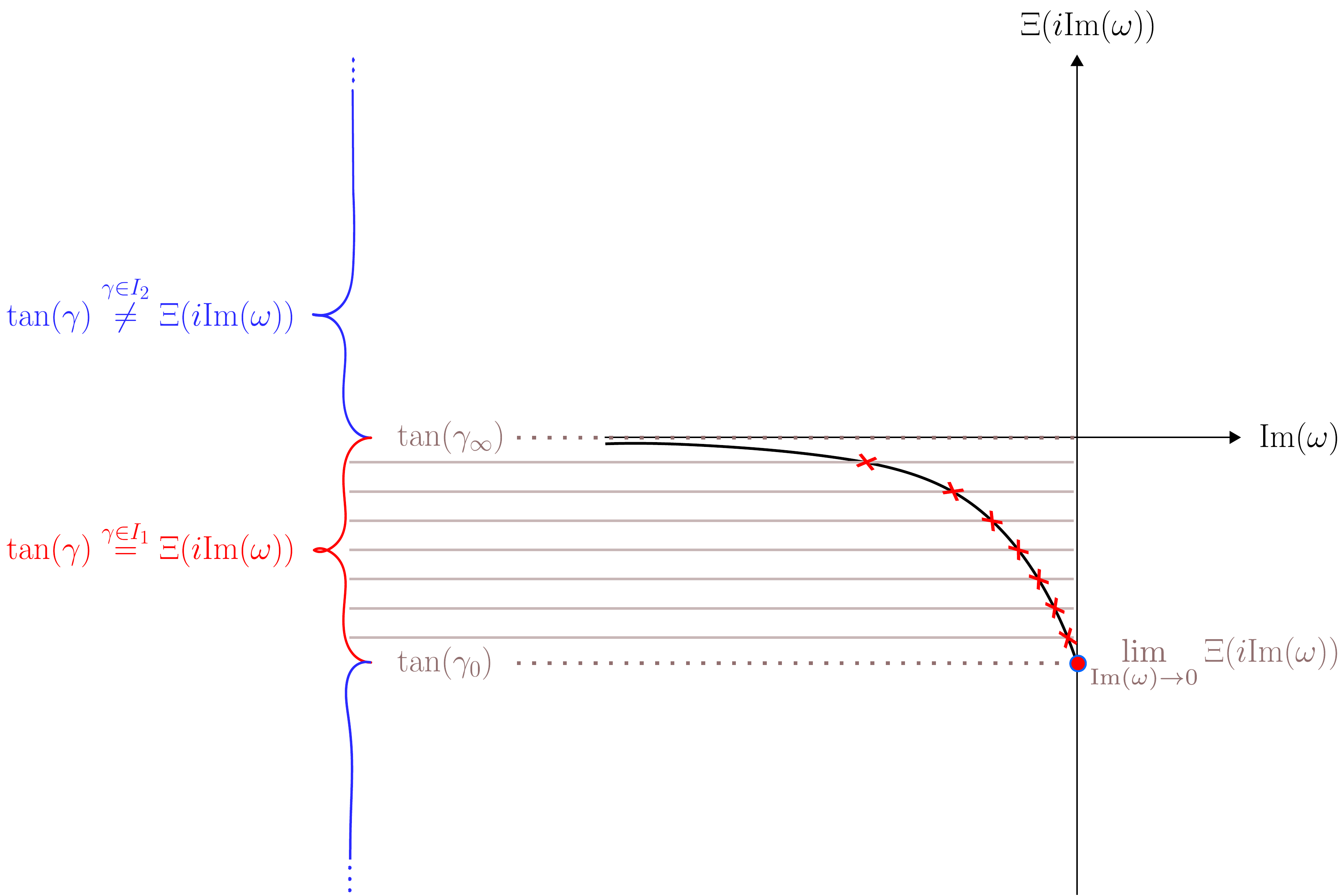}
 \caption{The behavior of $\Xi$ for purely imaginary frequencies with $\Imag(\omega)<0$ for $\nu\in(0,1)$.}
 \label{fig:Xi from 0 to 1}
 \end{figure}

 The behavior of $\Xi$ for general parameters, with $\nu\in(0,1)$, is illustrated in Figure \ref{fig:Xi from 0 to 1}. It has no poles, and as $\omega$ goes to zero, $\Xi$ goes to a $\lambda$-dependent limit \eqref{eq: limits of Xi k=0 at omega 0}. When $\Xi(\omega)$ lies in the interval $I_1$, $\chi(\omega)$ has one simple zero. When $\Xi(\omega)$ lies in the interval $I_2$, $\chi(\omega)$ has no zeros. This means that for each $\lambda$-mode, there is a different critical interval of Robin boundary conditions for which the radial Green function has poles, given by $\gamma\in[\gamma^\lambda_{c},\pi)$ with:
 \begin{equation}
   \gamma^\lambda_{c} := \arctan\left( \left(\frac{-\lambda L^2}{4}\right)^{-\nu}\frac{\Gamma(\nu)}{\Gamma(-\nu)} \right).
 \end{equation}
 However, we have that $\gamma^\lambda_{c}\in(\pi/2,\pi)$, which means that for all $\lambda$-modes, there are no bound states for $\gamma\in[0,\pi/2]$.
 \begin{figure}[H]
 \centering
   \includegraphics[width=.9\textwidth]{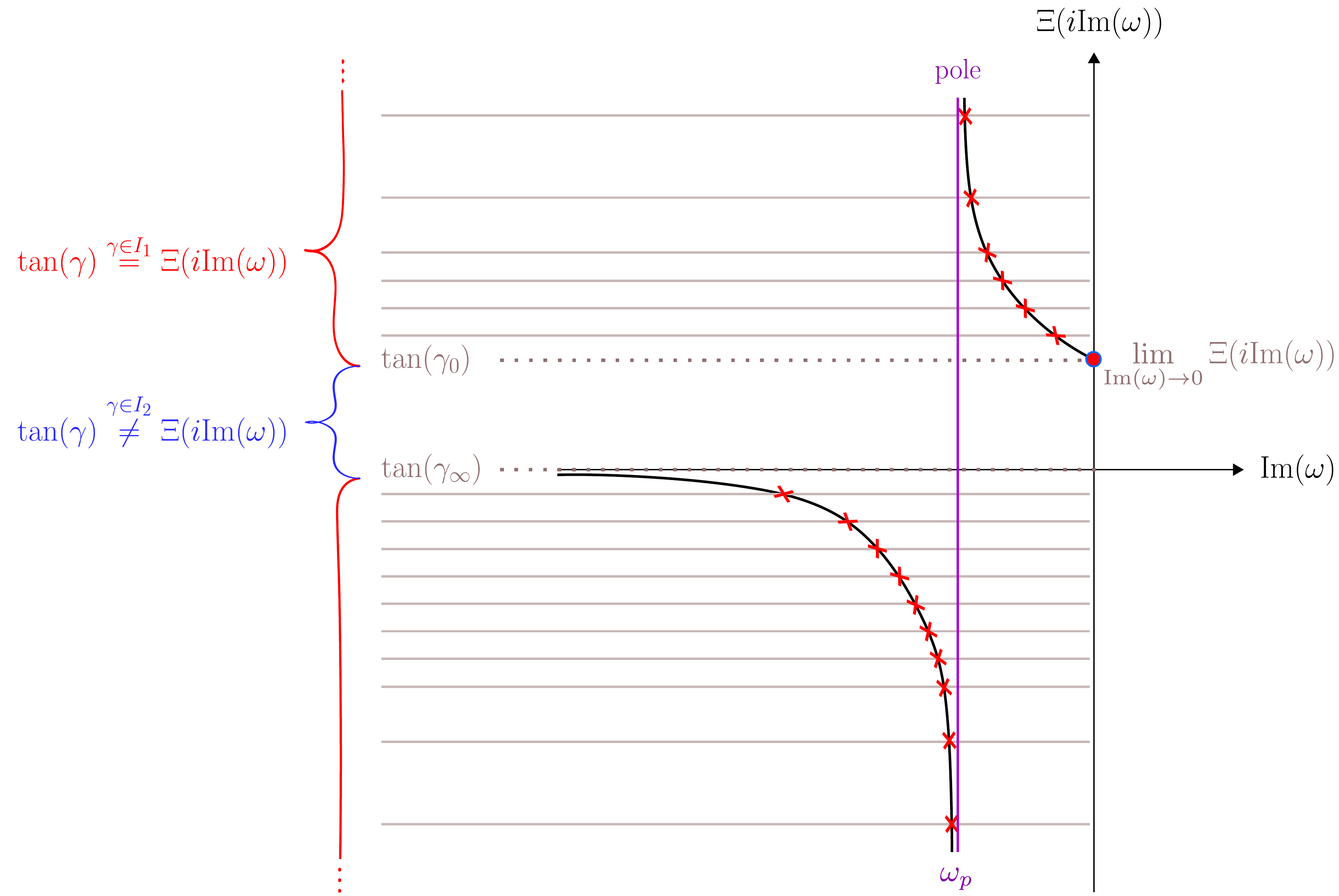}
 \caption{The behavior of $\Xi$ for purely imaginary frequencies with $\Imag(\omega)<0$ for $\nu\in(1,2)$.}
 \label{fig:Xi from 1 to 2}
 \end{figure}
 \noindent\underline{The analysis for $\nu\in(1,2)$:}\\

 The behavior of $\Xi$ for general parameters, with $\nu\in(1,2)$, is illustrated in Figure \ref{fig:Xi from 1 to 2}. In this case, $\Xi$ has one pole for each $\lambda$-mode. $\chi(\omega)$ has a simple zero when $\Xi$
lies in $I_1$, and no zeros when it lies in $I_2$. In this case, $I_2$ corresponds to boundary conditions $\gamma\in(0,\gamma^\lambda_{c})$.

 In short, for each $\lambda$-mode, the Robin boundary conditions for which there are no bound states are:
 \begin{equation}
   \label{eq: no bound states k 0 b.c.}
 \gamma\in(0,\gamma^\lambda_c) \quad \text{ with }\quad  \gamma^\lambda_c
\in \begin{cases}
                  (\pi/2,\pi) \, &\nu\in(0,1),\\
                 [0,\pi/2) \,     &\nu\in(1,2).
               \end{cases}
 \end{equation}

 We conclude that, if $\nu\in(0,1)$, then, for all $\lambda$-modes, the boundary conditions for which there are no bound states are  $\gamma\in[0,\pi/2]$. At the same time, if $\nu\in(1,2)$, for all boundary conditions $\gamma\in(0,\pi)$ there are $\lambda$-modes for which there are bound states, but Dirichlet and mode-dependent boundary conditions can be suitably imposed.

\section{The radial equation on the hyperbolic Lifshitz black hole}
\label{sec: The radial equation on the hyperbolic Lifshitz black hole, k-1}

In this section, we study the Klein-Gordon equation on $\mathsf{Lif}_{-1}$, an hyperbolic Lifshitz black hole. We follow the same procedure as in Section \ref{sec: the radial equation equation on Lifshitz spacetime, k=0}. First, in Section \ref{sec: The radial equation as a hypergeometric equation k=-1}, we show that the radial equation \eqref{eq: the radial equation} can be reduced to a hypergeometric equation under a suitable coordinate transform. In Section \ref{sec:The radial solutions k=-1}, we individuate the bases of radial solutions that are numerically satisfactory near each singularity. Then, in Section \ref{sec: Square-integrability conditions k=-1}, we study the conditions for the solutions to be square-integrable with respect to the measure \eqref{eq:measure mu(r)}.
With the symmetry properties described in Section \ref{sec: Symmetries under conjugation and reflection k -1}, we study the radial Green function defined in Section \ref{sec:The radial Green function k=-1}. Finally, in Section \ref{sec: on the existence of bound states for k=-1}, we verify for which boundary conditions the radial Green function has poles.

\subsection{The radial equation as a hypergeometric equation}
\label{sec: The radial equation as a hypergeometric equation k=-1}
  Applying the coordinate change
  \begin{equation}
    r\mapsto s = \frac{2 r^2 - L^2}{2r^2}\in (0,1),
  \end{equation}
  and making the ansatz
  \begin{align}
    \label{eq: ansatz R(s) k -1}
      &R(s) =  s^{-i L\omega} (1-s)^{\frac{1}{2} \left(2 + \nu \right)}
h(s),
  \end{align}
  we obtain that $R(s)$ solves the radial equation if and only if $h(s)$ is a solution of the hypergeometric equation
  \begin{subequations}
  \begin{align}
    \label{eq: hypergeometric h(s) k -1}
        s(1-s)h''(s) + (c-(a+b+1)s)h'(s) -abh(s)=0,
  \end{align}
  with parameters
  \begin{align}
    \label{eq: parameters of hypergeometric h(s) k -1}
      &a = \frac{1+\nu}{2} -iL\omega - \Upsilon,\\
      &b = \frac{1+\nu}{2} -iL\omega + \Upsilon,\\
      &c = 1-2iL\omega,
  \end{align}
  where
  \begin{align}
    \label{eq: parameter upsilon}
      &\Upsilon = - \frac{\sqrt{1 +2\lambda -4L^2\omega^2}}{2}  .
  \end{align}
\end{subequations}
\subsection{The radial solutions}
\label{sec:The radial solutions k=-1}

\noindent  When the hypergeometric parameters $$c,\, c-a-b, \text{ and }a-b$$ are not integers, numerically satisfactory bases of solutions of the
hypergeometric equation \eqref{eq: hypergeometric h(s) k -1}, dubbed $$\{h_{1(s_0)(s)},h_{2(s_0)}(s)\}$$ at each singular endpoint $s_0\in\{0,1\}$, are given in Table \ref{tab: basis hypergeo k -1}. They are respectively given by expressions (15.10.11)-(15.10.14) in \cite[Ch.15]{NIST}.

  \begin{table}[H]
    \centering
    \caption{Suitable bases of the hypergeometric equation \eqref{eq: hypergeometric h(s) k -1}.}\smallskip
    \label{tab: basis hypergeo k -1}
       \includegraphics[width=.8\textwidth]{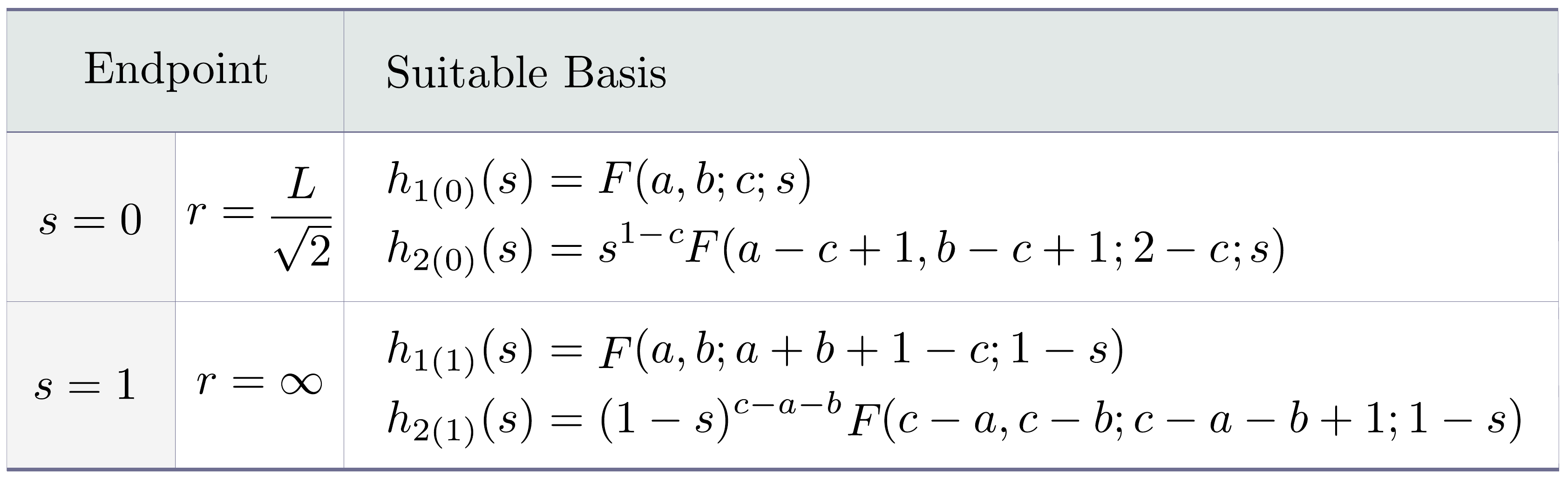}
   \end{table}
  A suitable basis at $s=s_0$ for the radial equation, dubbed $\{R_{1(s_0)}(s),R_{2(s_0)}(s)  \}$, descends from the ansatz \eqref{eq: ansatz R(s) k -1}; for $j\in\{1,2\}$
    \begin{align}
      \label{eq: basis R(s) k -1}
        &R_{j(s_0)}(s) =   s^{-i L\omega} (1-s)^{\frac{1}{2} \left(2 + \nu \right)}  h _{j(s_0)}(s).
    \end{align}

\subsection{Square-integrability conditions}
\label{sec: Square-integrability conditions k=-1}

\noindent Analogously to Section \ref{sec: Square-integrability conditions k=0}, we are interested at unveiling whether the solutions of the radial equation are square integrable close to the endpoints $r=\frac{L}{\sqrt{2}}$ and $r\to\infty$ with respect to the measure $q(r)dr$ where
$q$ is defined in Equation \eqref{eq:measure mu(r)}.  At the endpoint
$s=0$, $r=L/\sqrt{2}$, the radial solutions \eqref{eq: ansatz R(s) k -1}, behave as:
    \begin{subequations}
    \begin{align}
      &|R_{1(L/\sqrt{2})}(r)|^2 \sim  \left(2 r^2 - L^2\right)^{ + 2L \Imag(\omega)} ,\\
      &|R_{2(L/\sqrt{2})}(r)|^2 \sim   \left(2 r^2 - L^2\right)^{- 2L \Imag(\omega)}.
    \end{align}
  \end{subequations}
    With respect to the measure \eqref{eq:measure mu(r)}, we obtain that $R_{1(L/\sqrt{2})}(r)$ is square-integrable if and only if $\Imag(\omega)>0$, and $R_{2(L/\sqrt{2})}(r)$, if $\Imag(\omega)<0$. Thus  $r=L/\sqrt{2}$ is limit point and no boundary condition is necessary.

    At the endpoint $s=1$, $r\to\infty$, the asymptotic behavior of the
solutions is consistent with Equation \eqref{eq: asymptotic R(r) at radial infinity }:
    \begin{subequations}
    \begin{align}
      \label{eq:R asymp at infity k-1}
      &|R_{1(\infty)}(r)|^2 \sim  r^{-4 - 2\nu} ,\\
      &|R_{2(\infty)}(r)|^2 \sim  r^{-4 + 2\nu},
    \end{align}
    \end{subequations}
   Identically to the case $\kappa=0$ it descends that, for $\nu\in(0,1)\cup(1,2)$ both solutions are square integrable and $r=\infty$ is a limit circle. For $\nu=1$ or $\nu\geq2$, only $R_{1(\infty)}(r)$ is square-integrable and radial infinity is limit point.

    With the analyses above in mind, we denote the most general square-integrable radial solutions that are suitable at the singular endpoint $r=L/\sqrt{2}$ and $r=\infty$ by $R_{L/\sqrt{2}}(r)$ and $R_{\gamma}(r)$, respectively. Considering the solutions \eqref{eq: basis R(s) k -1} as functions of $r$ instead of $s$, they are given by
\begin{subequations}
  \label{final square integrable radial solutions k -1 function of r}
    \begin{align}
      \label{eq: square-integrable R(r) at r=L/sqrt2 k=-1}
      &R_{L/\sqrt{2}}(r) := \begin{cases}
                            R_{1(L/\sqrt{2})}(r), &\text{ if }\Imag(\omega)>0,\\
                            R_{2(L/\sqrt{2})}(r), &\text{ if }\Imag(\omega)<0,
                          \end{cases}
    \end{align}
    \begin{align}
      \label{eq: square-integrable R(r) at r=infty k=pm 1}
      &R_{\gamma}(r) := \cos(\gamma)R_{1(\infty)}(r) + \sin(\gamma)R_{2(\infty)}(r),
    \end{align}
    The solution $R_\gamma$ satisfies the generalized Robin boundary conditions at radial infinity parameterized by $\gamma\in[0,\pi)$, just as described in detail in Section \ref{sec: Robin boundary conditions at r=infty k 0} for the case $k=0$, but with $R_{1(\infty)}(r)$ and $R_{1(\infty)}(r)$ built out of \eqref{eq: basis R(s) k -1}.
\end{subequations}

\subsection{Symmetries under conjugation and reflection}
  \label{sec: Symmetries under conjugation and reflection k -1}

    The radial solutions for $\kappa=-1$ satisfy properties analogous to the ones described in Section \ref{sec: Symmetries under conjugation and reflection k=0}. The parameters of the hypergeometric equation are such that
    \begin{subequations}
      \label{eq:properties conjugation hypergeo k-1}
    \begin{align}
      & a(\overline{\omega}) = \overline{ a(\omega)-c(\omega)+1 },\\
      & b(\overline{\omega}) = \overline{ b(\omega)-c(\omega)+1 },\\
      & c(\overline{\omega}) = \overline{ 2-c(\omega) }.
    \end{align}
  \end{subequations}
    Using these properties, we find that for $j_1,j_2\in\{1,2\}$, the radial solutions \eqref{eq: basis R(s) k -1} satisfy
    \begin{equation}
      \label{eq: properties basis R(r) at r_0 k -1}
      R_{j_1(r_0)}(r,\overline{\omega}) = \overline{R_{j_2(r_0))}(r,\omega)}, \text{ with }\begin{cases}

        j_1 \neq j_2, &\text{ at } r_0=L/\sqrt{2},\\

        j_1 = j_2, &\text{ at } r_0=\infty.\\

                  \end{cases}
    \end{equation}
    Therefore, the radial solutions \eqref{final square integrable radial
solutions k -1 function of r} satisfy:
    \begin{subequations}
      \label{eq: properties square integrable R(r) at r_0 k -1}
    \begin{align}
      &R_{L/\sqrt{2}}(r,\overline{\omega}) =\overline{R_{L/\sqrt{2}}(r,\omega)} ,\\
      &R_{\gamma}(r,\overline{\omega}) =\overline{R_{\gamma}(r,\omega)}
.
    \end{align}
  \end{subequations}

\subsection{The radial Green function}
\label{sec:The radial Green function k=-1}
    Analogously to the case $\kappa=0$, described in Section \ref{sec: The radial Green function k 0}, we define the radial Green function $\mathcal{G}_{\omega}(r,r')$ for $\kappa=-1$, with radial solutions given as
in Equation \eqref{final square integrable radial solutions k -1 function
of r}.
    Considering the range of values $\nu \in (0, 1)\cup(1,2)$ and $p(r)$ as in Equation \eqref{eq:SL operator function p(r)}, the radial Green function reads
\begin{subequations}
       \begin{align}
         \label{eq: green function k -1}
       \mathcal{G}_{\omega}(r,r') =\frac{1}{ \mathcal{N}_\omega}\left( \Theta(r'-r) R_{L/\sqrt{2}}(r)R_{\gamma}(r') + \Theta(r-r')R_{L/\sqrt{2}}(r')R_{\gamma}(r)\right).
       \end{align}
       with normalization $ \mathcal{N}_\omega$
       \begin{equation}
         \label{eq: definition normalization of radial green function  k -1}
        \mathcal{N}_\omega := -p(r)W_r\left[ R_{L/\sqrt{2}}(r),R_{\gamma}(r)\right].
        \end{equation}
      \end{subequations}
Recall that $R_{L/\sqrt{2}}$ is defined by parts with respect to the sign
of the imaginary part of $\omega$.
When $\Imag(\omega)<0$, the fundamental relation connecting the square-integrable solution at $r=L/\sqrt{2}$ with the solutions at radial infinity is given by, \cite[(15.10.22)]{NIST}
\begin{subequations}
  \label{eq:fundamental k=-1 l/sqrt2 to inf}
\begin{equation}
  R_{2(L/\sqrt{2})}(r) = A_{\mini{-1}} R_{1(\infty)}(r) + B_{\mini{-1}}
R_{2(\infty)}(r),
\end{equation}
with coefficients
\begin{align}
  &A_{\mini{-1}}:= \frac{\Gamma(2-c)\Gamma(c-a-b)}{\Gamma(1-a)\Gamma(1-b)},\\
  &B_{\mini{-1}}:= \frac{\Gamma(2-c)\Gamma(a+b-c)}{\Gamma(a-c+1)\Gamma(b-c+1)}.
\end{align}
\end{subequations}
Using the fundamental relation \eqref{eq:fundamental k=-1 l/sqrt2 to inf}, and the Wronskian \cite[(15.10.3)]{NIST}, the normalization \eqref{eq: definition normalization of radial green function  k -1} can be written
as
  \begin{equation}
    \mathcal{N}_\omega=L^4\nu\{B_{\mini{-1}}\cos(\gamma) - A_{\mini{-1}}\sin(\gamma)\}, \text{ for }\Imag(\omega)<0.
  \end{equation}
Moreover, given the properties with respect to complex conjugation, as described in Section \ref{sec: Symmetries under conjugation and reflection k -1}, it holds:
\begin{equation}
  \mathcal{N}_{\overline{\omega}}=L^4\nu\{\overline{B_{\mini{-1}}}\cos(\gamma) - \overline{A_{\mini{-1}}}\sin(\gamma)\}, \text{ for }\Imag(\overline{\omega})>0.
\end{equation}
Analogously to the case $\kappa=0$:
\begin{equation}
  \label{eq: property normalization constant omega conjugate k-1}
\mathcal{N}_{\overline{\omega}}=\overline{\mathcal{N}_{\omega}} \quad\text{ and }\quad \mathcal{G}_{\overline{\omega}}(r,r')=\overline{\mathcal{G}_{\omega}(r,r')},
\end{equation}
which entails that we can analyze the case $\Imag(\omega)<0$, and extend directly the results to $\Imag(\omega)>0$. In the next section, we assume
$\Imag(\omega)<0$ and we investigate the zeros of the normalization.

\subsection{On the existence of bound states}
\label{sec: on the existence of bound states for k=-1}

In the following, we check the existence of bound states for each boundary condition by studying the zeros of the normalization $\mathcal{N}_\omega$. In all cases, consider $n\in\mathbb{N}_0$. Observe that, since we are
considering $\nu\in(0,1)\cup(1,2)$, the effective mass assumes only negative values and the coupling constant is positive as indicated in Equation
\eqref{eq: coupling for negative effective mass}. Accordingly, for $k=-1$, we have that the parameter $\lambda$, given by Equation \eqref{eq: lambda shited by xi}, is always negative.

\subsubsection{Dirichlet boundary condition, $\gamma=0$}
\vspace{-1cm}
\begin{align}
  \mathcal{N}_{\omega} =0 &\iff
   \left(a+2iL\omega = -n \text{ or }b+2iL\omega = -n\right) \nonumber \\
    & \iff \omega = \frac{i \left(-2 \lambda +\nu ^2+2 \nu +4 n^2+4 \nu
 n+4 n\right)}{4 L (\nu +2 n+1)}.
\end{align}
These frequencies have positive imaginary part, and the expression above for the normalization is defined for $\Imag(\omega)<0$, hence there are no zeros in its domain.
\subsubsection{Neumann boundary condition, $\gamma=\pi/2$}
\begin{figure}[H]
  \centering
  \includegraphics[align=c,width=.4\textwidth]{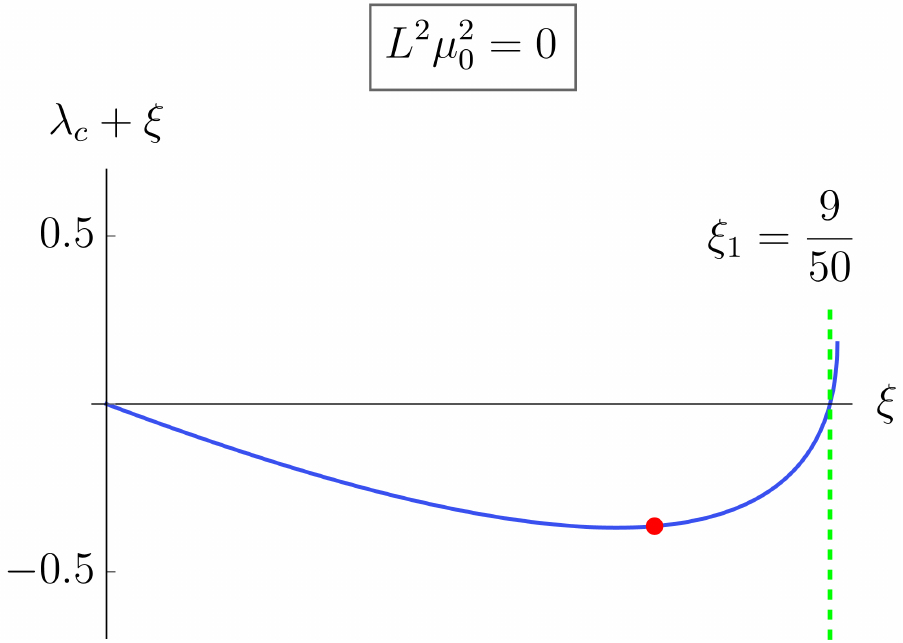}\hspace{.5cm}%
  \includegraphics[align=c,width=.4\textwidth]{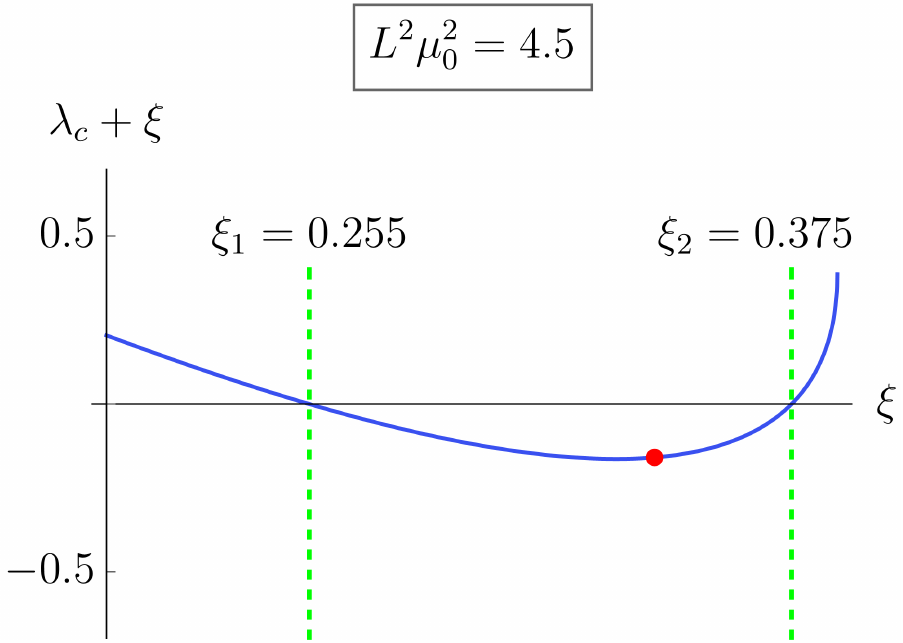}\vspace{.7cm}
  \includegraphics[align=c,width=.4\textwidth]{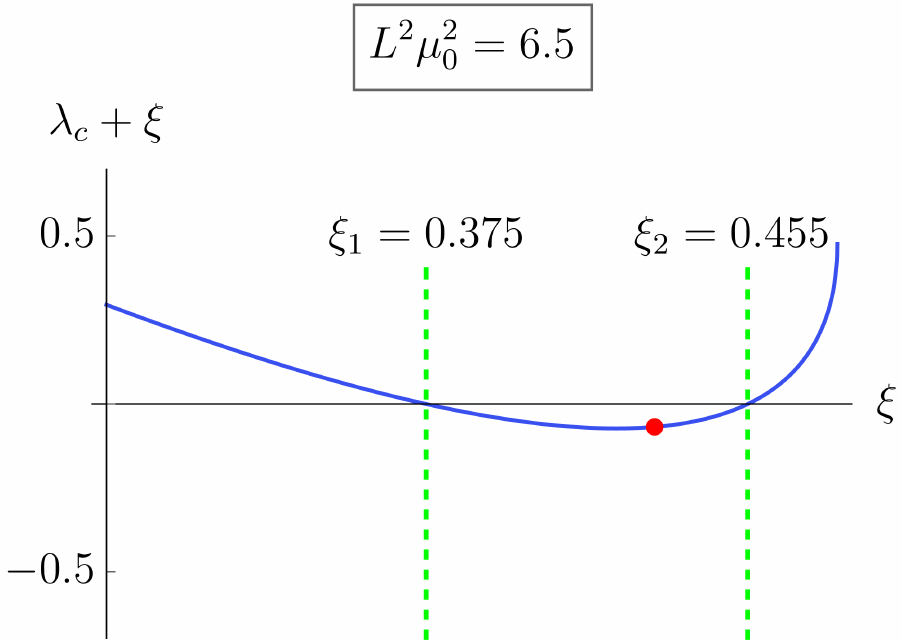}\hspace{.5cm}%
   \includegraphics[align=c,width=.4\textwidth]{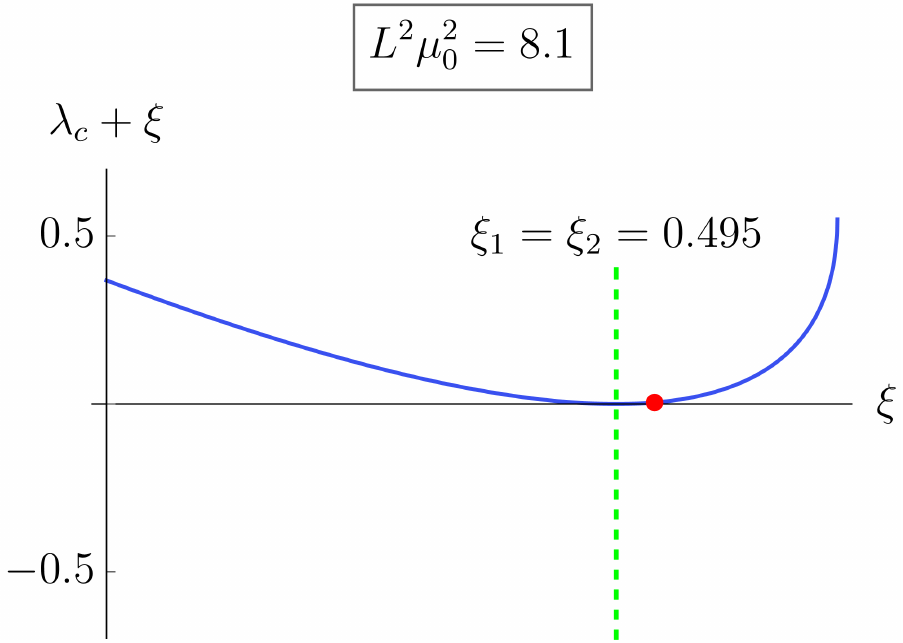}
    \caption{$\lambda_c + \xi$ as a function of $\xi\in\left(\frac{L^2\mu_0^2}{22},\frac{L^2\mu_0^2+4}{22}\right)$ for different masses. To the left of the (red) point, $\nu\in(1,2)$; to the right, $\nu\in(0,1)$.}
  \label{fig:lambda critical function of coupling massless case k = -1}
  \end{figure}

\vspace{-1cm}
\begin{align}
  \mathcal{N}_{\omega} =0 &\iff
   \left(1-a = -n \text{ or }1-b = -n\right) \nonumber \\
    & \iff \omega = \frac{i \left(-2 \lambda +\nu ^2-2 \nu +4 n^2-4 \nu
 n+4 n\right)}{4 L (-\nu +2 n+1)}.
\end{align}
For $n>0$,  the frequency above lies in the upper part of the complex-plane. For $n=0$, however, its imaginary part does assume negative values.
Let $\omega_p$ be the frequency for $n=0$:
\begin{align}
  \label{eq: zero neumann k -1}
  \omega_p =   \frac{i(-2\lambda + \nu(\nu-2))}{4L(1-\nu)}.
\end{align}
and let us define the auxiliary quantities:
\begin{equation}
  \label{eq: lambda c neumann km1}
  \lambda_c:=-\frac{\nu(2-\nu)}{2},
\end{equation}
and
\begin{align}
  \label{eq: xi1 and xi2 neumann km1}
                \xi_{1} :=   \frac{1}{100} \left(9+ 5 L^2 \mu_0^2 - \sqrt{81-10 L^2 \mu_0^2}\right) ,\\
                \xi_{2} :=   \frac{1}{100} \left(9+ 5 L^2 \mu_0^2 + \sqrt{81-10 L^2 \mu_0^2}\right) .
\end{align}
For the frequency $\omega_p$, we find that
\begin{align}
  \label{eq: zero neumann k -1 condition negative Im}
  \Imag(\omega_p) <0  \text{ for } \begin{cases}
                                      \lambda>\lambda_c, \text{ if }\nu\in(0,1), \\
                                      \lambda<\lambda_c, \text{ if }\nu\in(1,2).
                                    \end{cases}
\end{align}
It is easy to see that in the range $\nu\in(1,2)$, regardless of the value of $\lambda_c$, there is an uncountable set of $\lambda$-modes for which $\lambda<\lambda_c$. For $\nu\in(0,1)$, on the other hand, Equation \eqref{eq: zero neumann k -1 condition negative Im} is meaningful only if $\lambda_c < -\xi$. It follows that:
\begin{equation}
  \lambda_c \geq -\xi \iff\text { either}
\begin{cases}
  \text{(i) } &\hspace{-0.5cm}  L^2\mu_0^2 < 8.1 \text{ and }(\xi\leq\xi_1 \text{ or }\xi\geq\xi_2);\\
  \text{(ii) } &\hspace{-0.5cm} L^2\mu_0^2 \geq  8.1.\\
  \end{cases}
\end{equation}
For the massless field, $\xi_1=0$ and $\xi_2=\frac{9}{50}$. Therefore, $\lambda_c \geq -\xi$ if $\xi\in\left[\frac{9}{50},\frac{4}{22}\right)$. For massive fields, the behaviour of $\lambda_c$ is analogous to the one for $\mu_0=0$, but for increasing values of $L^2\mu_0^2$, $\lambda_c +\xi$ as a function of $\xi$ is shifted upwards, as illutrated in Figure \ref{fig:lambda critical function of coupling massless case k = -1}.

\noindent All included, the values of $L$, $\mu_0$ and $\xi$ for which no
bound states emerge are:
\begin{subequations}
  \label{eq: no poles km1 neumann}
\begin{align}
    &\nu\in(0,1) \begin{cases}
    \text{(i) } \,\,L^2\mu_0^2 < 8.1, \xi\in \left(\xi_1,\xi_2\right) \text{ and }\lambda\leq\lambda_c; \\
    \text{(ii) }\, L^2\mu_0^2 < 8.1 \text{ and }\xi\notin \left(\xi_1,\xi_2\right) ; \\
    \text{(iii) } L^2\mu_0^2 \geq 8.1; \end{cases}\\
    &\nu\in(1,2)\begin{cases}
    \text{(iv) } L^2\mu_0^2 < 8.1,\, \xi \in \left(\xi_1,\xi_2\right) \text{ and }\lambda\geq\lambda_c; \\
    \text{(v) }\,L^2\mu_0^2 = 8.1, \, \xi=\xi_1=\xi_2 \text{ and }\lambda=-\xi.\end{cases}
    \end{align}
\end{subequations}
Observe that for $\nu\in(0,1)$, conditions (ii) and (iii) holds for all $\lambda$-modes, while the other conditions are  mode-dependent.
\subsubsection{Robin boundary conditions, $\gamma\in(0,\pi/2)\cup(\pi/2,\pi)$}
\noindent Define
\begin{subequations}
\begin{align}
  &\zeta(\omega) := \tan(\gamma) - \eta(\omega),\label{eq:zeta = tan - kappa} \\
  &\eta(\omega) := \frac{\Gamma(a+b-c)}{\Gamma(c-a-b)}\frac{\Gamma(1-a)\Gamma(1-b)}{\Gamma(a-c+1)\Gamma(b-c+1)}.\label{eq: kappa definition}
\end{align}
\end{subequations}
The bound states are the frequencies for which $\zeta(\omega)=0$, which
are in turn those for which $\eta(\omega)\in\mathbb{R}$. Hence, let us study the function $\eta$. First, we note that, as in the case $\kappa=0$, $\eta(\omega)\in\mathbb{R} \iff \omega = -i |\Imag(\omega)|$. In fact, the lines of $\Imag(\kappa)=0$ are very similar to those of Figure \ref{fig:Im(Xi) on the complex plane k = 0}. Moreover,
\begin{subequations}
\begin{align}
  &\lim\limits_{\omega\rightarrow 0} \eta( \omega) =  \frac{\Gamma(\nu)}{\Gamma(-\nu)}\frac{\Gamma\left( \frac{1-\nu}{2} -\frac{\sqrt{1+2\lambda}}{2} \right)\Gamma\left( \frac{1-\nu}{2} +\frac{\sqrt{1+2\lambda}}{2} \right)}{\Gamma\left( \frac{1+\nu}{2}  +\frac{\sqrt{1+2\lambda}}{2}\right)\Gamma\left( \frac{1+\nu}{2}  -\frac{\sqrt{1+2\lambda}}{2} \right)},\label{eq: limits of kappa k=-1 at 0}\\
  &\lim\limits_{\omega\rightarrow \omega_p} \eta(\omega) = \infty , \label{eq: limits of kappa k=-1 at omegap}\\
  &\lim\limits_{\omega\rightarrow \infty} \eta(- i \omega) = 0 .\label{eq: limits of kappa k=-1 at infty}
\end{align}
\end{subequations}
 Note that the limit $\eta(0)$ is $\lambda$-dependent and that, whenever $\lambda$ can assume a critical value  $\lambda_c=-\frac{\nu(2-\nu)}{2}$, then $\omega_p=0$ and $\eta(0)\to\infty$. For $\lambda\neq \lambda_c$, $\eta(0)$ is a finite number that is negative for $\{\lambda<\lambda_c$ and $\nu\in(0,1)\}$, or for $\{\lambda>\lambda_c$ and $\nu\in(1,2)\}$, and positive otherwise.
As a consequence the function $\eta(i\Imag(\omega))$ with $\Imag(\omega)<0$ behaves like the function $\Xi(i\Imag(\omega))$, as in Figure \ref{fig:Xi from 0 to 1}, whenever there are no bound states. Since, in this case, $\lambda$ is never equal to $\lambda_c$, the acceptable boundary conditions are such that
\begin{equation}
  \gamma\in \left[0, \gamma_{c}^\lambda \right) \quad \text{ with }\quad
\gamma_{c}^\lambda := \arctan\left( \eta( 0) \right)\in\left(\frac{\pi}{2},\pi\right).
\end{equation}
Whenever there are bound states, $\eta(i\Imag(\omega))$ behaves as $\Xi(i\Imag(\omega))$ in Figure \ref{fig:Xi from 1 to 2}. In this case, for each $\lambda$-mode, there are no bound states in the regime
\begin{equation}
  \label{eq: no bound states k -1 b.c.}
\gamma\in[0,\gamma^\lambda_{c})\text{ with }   \gamma^\lambda_{c} \in (0,\pi/2).
\end{equation}

\section{The radial equation on the spherical Lifshitz black hole}
\label{sec: The radial equation on the spherical Lifshitz black hole, k+1}

In this section, we study the Klein-Gordon equation on $\mathsf{Lif}_1$, a spherical Lifshitz black hole. We follow the same steps as in the previous two cases. First, in Section \ref{sec: The radial equation as a hypergeometric equation k=+1}, we write the radial equation as an hypergeometric equation, and, in Section \ref{sec: The radial solutions k+1}, we state the bases that are numerically satisfactory near each singular endpoint.
Then, after checking, in Section \ref{sec: Square-integrability conditions k+1}, which solutions are compatible with self-adjoint extensions of the radial part of the Klein-Gordon operator, we describe, in Section \ref{sec: Symmetries under conjugation and reflection k +1}, the symmetry properties satisfied by the solutions. Finally, in Section \ref{sec: The radial Green function k+1}, we define the radial Green function and we verify, in Section \ref{sec: On the existence of bound states for k=+1}, which boundary conditions can be imposed bypassing the emergence of bound states.

Observe that, even though the notation used in the following is the same as the one of Section \ref{sec: The radial equation on the hyperbolic Lifshitz black hole, k-1} , the parameters and functions have different definitions here.

  \subsection{The radial equation as a hypergeometric equation}
  \label{sec: The radial equation as a hypergeometric equation k=+1}
\noindent  Applying the coordinate change:
  \begin{equation}
    r\mapsto s = \frac{2 r^2 + L^2}{2r^2}\in(1,\infty),
  \end{equation}
  and considering the ansatz
  \begin{align}
    \label{eq: ansatz R(s) k +1}
      &R(s) =  s^{i L\omega} (s-1)^{\frac{1}{2} \left(2 + \nu \right)} h(s),
  \end{align}
  we obtain that \eqref{eq: ansatz R(s) k +1} solves the radial equation if and only if $h(s)$ is in turn a solution of the hypergeometric equation
  \begin{subequations}
  \begin{align}
    \label{eq: hypergeometric h(s) k +1}
        s(1-s)h''(s) + (c-(a+b+1)s)h'(s) -abh(s)=0,
  \end{align}
  with parameters
  \begin{align}
    \label{eq: parameters of hypergeometric h(s) k +1}
      &a = \frac{1+\nu}{2} + iL\omega- \Upsilon,\\
      &b = \frac{1+\nu}{2} + iL\omega + \Upsilon,\\
      &c = 1+2iL\omega,
  \end{align}
where
  \begin{align}
    \label{eq: parameter upsilon}
      &\Upsilon = \frac{\sqrt{1 -2\lambda -4L^2\omega^2}}{2}  .
  \end{align}
\end{subequations}

\subsection{The radial solutions}
\label{sec: The radial solutions k+1}
A basis for the radial equation, suitable at each singular endpoint $s=s_0\in\{1,\infty\}$, is given by $\{R_{1(s_0)}(s),R_{2(s_0)}(s) \}$, with
  \begin{align}
    \label{eq: basis R(s) k +1}
      &R_{j(s_0)}(s) =  s^{i L\omega} (s-1)^{\frac{1}{2} \left(2 + \nu \right)}  h _{j(s_0)}(s), \text{ for }j\in\{1,2\},
  \end{align}
 and where $\{h_{1(s_0)(s)},h_{2(s_0)}(s)\}$ is a basis of solutions, at $s_0$, of the hypergeometric equation \eqref{eq: hypergeometric h(s) k +1}. When the hypergeometric parameters $$c,\, c-a-b, \text{ and }a-b$$ are
not integers, we consider the solutions given in Table \ref{tab: basis hypergeo +1}. Note that, since $s>1$, we have conveniently redefined the bases at $r=\infty$, in comparison with the $\kappa=0$ case that is given in Table \ref{tab: basis confluent hypergeo}. The reason is that it places the branch cut of the solution $h_{2(0)}$ outside of the domain of the equation. When the above mentioned hypergeometric parameters assume integer values, another secondary solution must be chosen. As in the case $\kappa=-1$, such solutions are not square-integrable. Thus, even though
we consider them in the analysis, we do not include their explicit expressions here.
 This choice of basis grants us analogous symmetry properties with respect to complex conjugation, as we describe in the following Section \ref{sec: Symmetries under conjugation and reflection k +1}.

 \setlength\extrarowheight{6pt} 
  \begin{table}[H]
    \centering
    \caption{Suitable basis of the hypergeometric equation \eqref{eq: hypergeometric h(s) k +1}.}\smallskip
    \label{tab: basis hypergeo +1}
       \includegraphics[width=.8\textwidth]{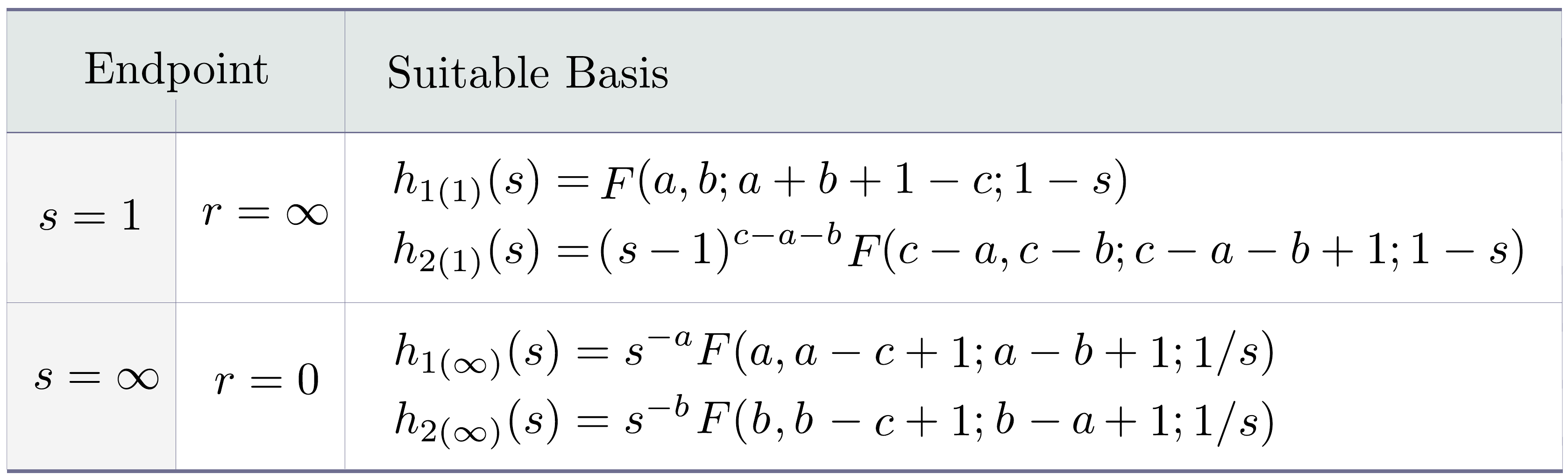}
   \end{table}
\subsection{Square-integrability conditions}
\label{sec: Square-integrability conditions k+1}

    The asymptotic behavior of the solutions at the endpoint $s=1$, $r=\infty$ is consistent with \eqref{eq: asymptotic R(r) at radial infinity }, and exactly as stated in the $\kappa=-1$ case in Equation \eqref{eq:R asymp at infity k-1}. In brief, for $\nu\in(0,1)\cup(1,2)$ both solutions are square integrable and $r=\infty$ is a limit circle. For $\nu=1$ or $\nu\geq2$, only $R_{1(\infty)}(r)$ is square-integrable and radial infinity is limit point.

    At $s=\infty$, $r=0$, we have
    \begin{subequations}
    \begin{align}
      &|R_{1(0)}(r)|^2 \sim  r^{-2-4\Real(\Upsilon)} ,\\
      &|R_{2(0)}(r)|^2 \sim  r^{-2+4\Real(\Upsilon)},
    \end{align}
  \end{subequations}
    and since $q(r)\sim r$, $r=0$ is limit point. When $\Real(\Upsilon)<0$, only $R_{1(0)}(r)$ is square-integrable, while if $\Real(\Upsilon)>0$, then only $R_{2(0)}(r)$ is square-integrable. However, directly from the definition of $\Upsilon$, given by \eqref{eq: parameter upsilon}, it follows that $\Real(\Upsilon)$ is always non-negative if we choose the
principal root.

    We denote the most general square-integrable radial solutions that are suitable at singular endpoint $r=0$ and $r=\infty$, respectively by
$R_{0}(r)$ and $R_{\gamma}(r)$, defined by
\begin{subequations}
  \label{final square integrable radial solutions k +1 function of r}
    \begin{align}
      \label{eq: square-integrable R(r) at r=0 k=-1}
      &R_{0}(r) := R_{2(0)}(r), \text{ if }\Real(\Upsilon)> 0.
    \end{align}
    \begin{align}
      \label{eq: square-integrable R(r) at r=infty k=pm 1}
      &R_{\gamma}(r) := \cos(\gamma)R_{1(\infty)}(r) + \sin(\gamma)R_{2(\infty)}(r),
    \end{align}
    The solution $R_\gamma$ satisfies the generalized Robin boundary conditions at radial infinity parameterized by $\gamma\in[0,\pi)$, as for the
cases $\kappa=0$ and $\kappa=-1$.
\end{subequations}

\subsection{Symmetries under conjugation and reflection}
\label{sec: Symmetries under conjugation and reflection k +1}

    The parameters of the hypergeometric equation satisfy the same properties of the case $\kappa=-1$, see Equation \eqref{eq:properties conjugation hypergeo k-1}. As a consequence, the radial solutions satisfy
    \begin{align}
      \label{eq: properties basis R(r) at r_0 k +1}
      &R_{j(r_0)}(r,\overline{\omega}) = \overline{R_{j(r_0)}(r,\omega)}, \text{ for }j\in\{1,2\}, \text{ and } r_0\in\{0,\infty\}.
      \end{align}
    Accordingly, we have for the general square-integrable solutions \eqref{final square integrable radial solutions k +1 function of r}
    \begin{align}
      \label{eq: symm prop sol sq int k+1}
        R_{0}(r,\overline{\omega})= \overline{R_{0}(r,\omega)},\\
        R_{\gamma}(r,\overline{\omega})= \overline{R_{\gamma}(r,\omega)} .
    \end{align}

    \subsection{The radial Green function}
    \label{sec: The radial Green function k+1}
    Considering the range of values $\nu \in (0, 1)\cup(1,2)$ and $p(r)$ as in \eqref{eq:SL operator function p(r)}, the radial Green function $\mathcal{G}_{\omega}(r,r')$ for $\kappa=+1$ is written in terms of the radial solutions given by Equation \eqref{eq: square-integrable R(r) at r=0 k=-1} and \eqref{eq: square-integrable R(r) at r=infty k=pm 1}, as
       \begin{align}
         \label{eq: green function k +1}
       \mathcal{G}_{\omega}(r,r') =\frac{1}{ \mathcal{N}_\omega}\left( \Theta(r'-r) R_{0}(r)R_{\gamma}(r') + \Theta(r-r')R_{0}(r')R_{\gamma}(r)\right),
       \end{align}
       with normalization
       \begin{equation}
         \label{eq: definition normalization of radial green function  k +1}
        \mathcal{N}_\omega := -p(r)W_r\left[ R_{0}(r),R_{\gamma}(r)\right].
        \end{equation}
The Green function and its normalization are defined for all frequencies such that $\Real(\Upsilon)>0$. If we always choose the principle root of a complex number, then, $\Real(\Upsilon)$ is never negative, and it holds
that:
\begin{equation}
  \label{eq: real upsilon = 0 omega interval}
  \Real(\Upsilon)=0  \iff \begin{cases}
  \omega = \Re(\omega) \in\left(\infty,-\omega_c\right]\cup\left[+\omega_c,\infty\right) &\text{ and }\lambda < \frac{1}{2}, \\
  \omega = \Re(\omega)  &\text{ and }\lambda =  \frac{1}{2}, \\
  \omega = \Re(\omega)\neq 0  &\text{ and }\lambda > \frac{1}{2}, \\
  \omega = i\Imag(\omega)\in\left[-\omega_c , +\omega_c\right] &\text{ and }\lambda > \frac{1}{2},
\end{cases}
\end{equation}
where the parameter $\omega_c$ is defined as:
\begin{equation}
  \label{eq: limiting value of real frequency k=+1}
  \omega_c = \frac{\sqrt{| 1-2\lambda| }}{2L}.
\end{equation}
Since $\lambda=\lambda_\ell^m + \xi\in(-\infty,\xi)$, the conditions in
Equation \eqref{eq: real upsilon = 0 omega interval} imply that $\xi\geq\frac{1}{2}$ gives rise to a $\lambda$-mode dependent domain for $\mathcal{N}_\omega$, which also entails $\lambda$-mode dependent boundary conditions. In this case, the existence of bound states depends crucially on the values of $\mu_0$ and $\xi$, as it did on the hyperbolic Lifshitz black hole. A preliminary analysis showed that for $\xi>1$, bound states emerge even for Dirichlet boundary conditions. For simplicity, in the following, we expose a detailed study considering $\xi<\frac{1}{2}$, which restricts the mass of the field in agreement with Equation \eqref{eq: coupling
for negative effective mass}. Accordingly, $D(\mathcal{N}_\omega)$, the domain of $\mathcal{N}_\omega$ reads
\begin{equation}
  \label{eq: domain of N kp1}
  D(\mathcal{N}_\omega)= \mathbb{C}\setminus \{\omega\;|\; |\textrm{Re}(\omega)|\geq\omega_c\;\textrm{and}\;\textrm{Im}(\omega)=0\}.
\end{equation}

In order to give an explicit expression of $\mathcal{N}_\omega$, observe that a square-integrable solution for $\Real(\Upsilon) > 0$ can be written as a linear combination of the solutions at radial infinity by \cite[(15.10.24)]{NIST}
\begin{align}
  \label{eq: connection R10 and R20 with R1 and 2 at infty}
  &R_{2(0)}(r) =A_{\mini{+1}} R_{1(\infty)}(r) + B_{\mini{+1}} R_{2(\infty)}(r).
\end{align}
with
\begin{align}
  &A_{\mini{+1}}:= \frac{\Gamma(b-a+1)\Gamma(c-a-b)}{\Gamma(1-a)\Gamma(c-a)},\\
  &B_{\mini{+1}}:= \frac{\Gamma(b-a+1)\Gamma(a+b-c)}{\Gamma(b)\Gamma(b-c+1)}.
\end{align}
Using Equation \eqref{eq: connection R10 and R20 with R1 and 2 at infty} and the Wronskian \cite[(15.10.5)]{NIST}, the normalization \eqref{eq: definition normalization of radial green function  k +1} reads
\begin{align}
  \mathcal{N}_{\omega} & = L^4\nu \left\{B_{\mini{+1}} \cos(\gamma)-A_{\mini{+1}}  \sin(\gamma)\right\}.
\end{align}
Since the coefficients above are such that $A_{\mini{+1}}(\overline{\omega})=\overline{ A_{\mini{+1}}(\omega)}$ and $B_{\mini{+1}}(\overline{\omega})=\overline{ B_{\mini{+1}}(\omega)}$, we also have
\begin{equation}
  \mathcal{N}_{\overline{\omega}}=\overline{\mathcal{N}_{\omega}} \quad \text{ and }\quad \mathcal{G}_{\overline{\omega}}=\overline{\mathcal{G}_{\omega}} .
\end{equation}

\subsection{On the existence of bound states}
\label{sec: On the existence of bound states for k=+1}

In this section, we study the existence of bound states for $\xi<\frac{1}{2}$ and $\Real(\Upsilon) > 0$, which yields \eqref{eq: domain of N kp1}.
 We consider the cases of Dirichlet, Neumann and Robin boundary conditions separately, taking $n\in\mathbb{N}_0$.
\subsubsection{Dirichlet boundary condition, $\gamma=0$}
\vspace{-1cm}
\begin{align}
  \mathcal{N}_{\omega} =0&\iff \{b = -n \text{ or } b-c+1\} \nonumber
\\
  &\iff \omega =  \pm \frac{i \left(2 \lambda +\nu ^2+2 \nu +4 n^2+4 \nu  n+4 n\right)}{4 L (\nu +2 n+1)} \text{ and } \lambda\geq \lambda_{n},
\end{align}
where
\begin{align}
  &\lambda_n:=\frac{2+4n +4n^2+2\nu+4n\nu+\nu^2}{2}.
\end{align}

Since $\lambda=\lambda_\ell^m + \xi \leq\xi<\frac{1}{2}$, and clearly $\lambda_n>1$, the frequencies above do not represent poles of the radial Green function. Yet, it is easy to see that if $\xi>1$, then poles would emerge even for Dirichlet boundary condition, depending on the value of the other parameters.

\subsubsection{Neumann boundary condition, $\gamma=\pi/2$}
\begin{figure}[H]
  \centering
   \includegraphics[align=c,width=.45\textwidth]{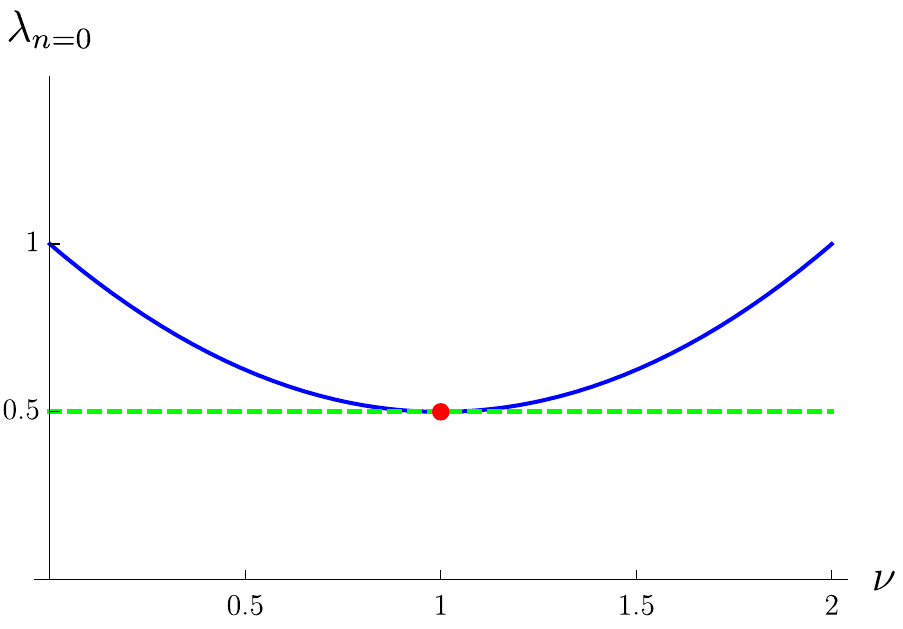}\hspace{.5cm}%
    \caption{$\lambda_{n=0}$ as a function of $\nu$. The (red) dot has coordinates $\left(1,0.5\right)$. }
  \label{fig: lambda n=0 neumann kp1}
  \end{figure}

\vspace{-1cm}
\begin{align}
  \mathcal{N}_{\omega} =0 &\iff
   \left(1-a = -n \text{ or }c-a = -n\right).
\end{align}
Defining the auxiliary quantities
\begin{align}
  &\omega_n := \frac{i \left(2 \lambda +\nu ^2-2 \nu +4 n^2-4 \nu  n+4 n\right)}{4 L (-\nu +2 n+1)},\\
  &\lambda_n:=\frac{2+4n +4n^2-2\nu-4n\nu+\nu^2}{2},
\end{align}
we find that
\begin{align}
    &1-a = 0  \iff  \omega = - \omega_{n=0} \text{ and }  \begin{cases}
      \lambda\geq \lambda_{n=0}, &\text{ for }\nu\in(0,1),\\
      \lambda\leq \lambda_{n=0}, &\text{ for }\nu\in(1,2).
  \end{cases}
  \end{align}
  \begin{align}
   &c-a = 0  \iff \omega = + \omega_{n=0}  \text{ and }  \begin{cases}
     \lambda\geq \lambda_{n=0}, &\text{ for }\nu\in(0,1),\\
     \lambda\leq \lambda_{n=0}, &\text{ for }\nu\in(1,2).
 \end{cases}
\end{align}
\begin{align}
   &1-a = -n \neq 0 \iff \omega = - \omega_n \text{ and } \lambda \geq \lambda_{n} ,\\
   &c-a = -n \neq 0 \iff \omega = +  \omega_n \text{ and } \lambda \geq \lambda_{n} .
\end{align}
For $\nu\in(0,1)\cup(1,2)$, it holds that $\lambda_{n=0}>\frac{1}{2}$, as illustrated in Figure \ref{fig: lambda n=0 neumann kp1}. Moreover, $\lambda_{n}$ is an increasing function of $n$ since $\frac{d\lambda_n}{dn}\big|_{n>0} = (2-2 \nu +4 n)|_{n>0}>0$. We conclude that for $n=0$, the normalization has no zeros for $\nu\in(0,1)$, and has two zeros at $\omega=\pm\omega_{n=0}$ for $\nu\in(1,2)$. On the other hand, for $n>0$, there are no zeros for any value of $\nu$.

\subsubsection{Robin boundary conditions, $\gamma\in(0,\pi/2)\cup(\pi/2,\pi)$}
\begin{figure}[H]
  \centering
   \includegraphics[align=c,width=.4\textwidth]{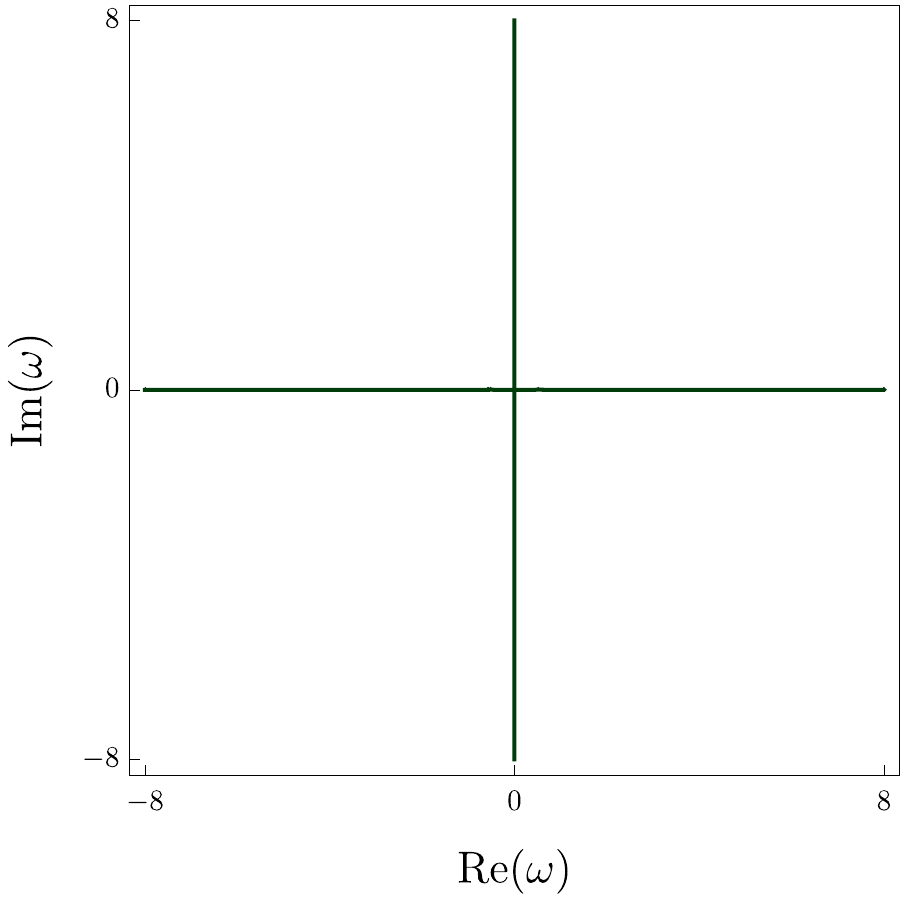}
  \caption{Contour lines of $\Imag(\eta)=0$ for $\mu_0=0$, $L=1$, $\lambda_\ell^m=0$ and $\xi=\frac{7}{88}$.}
  \label{fig:kappa on the complex plane k=+1}
    \end{figure}

\begin{figure}[H]
  \centering
  \includegraphics[align=c,width=\textwidth]{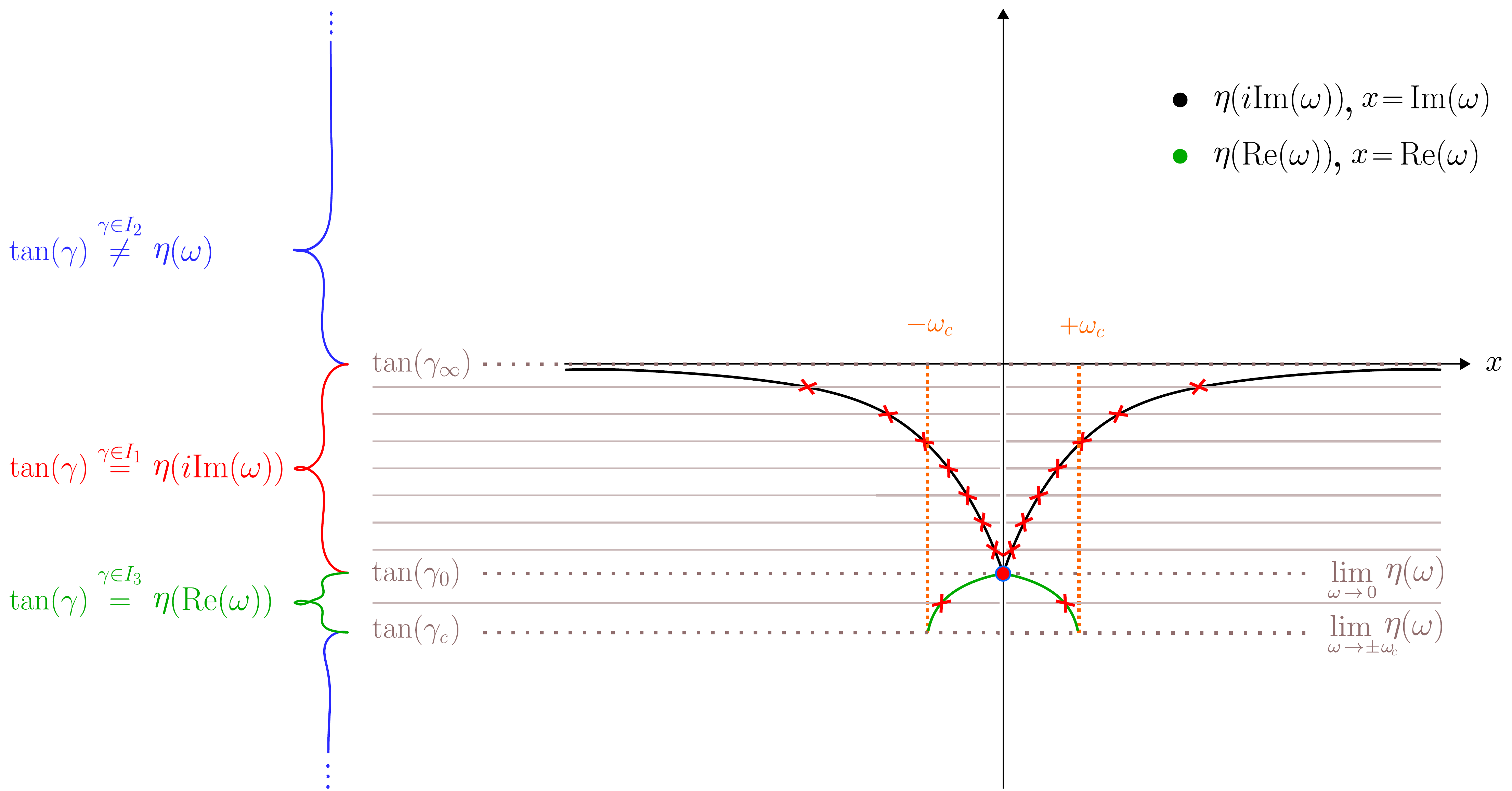}
    \caption{The behavior of $\eta$ for frequencies such that $\omega^2\in\mathbb{R}$ and $\nu\in(0,1)$.}
    \label{fig:kappa from 0 to 1 k+1}
  \end{figure}

\begin{figure}[H]
\centering
  \includegraphics[width=\textwidth]{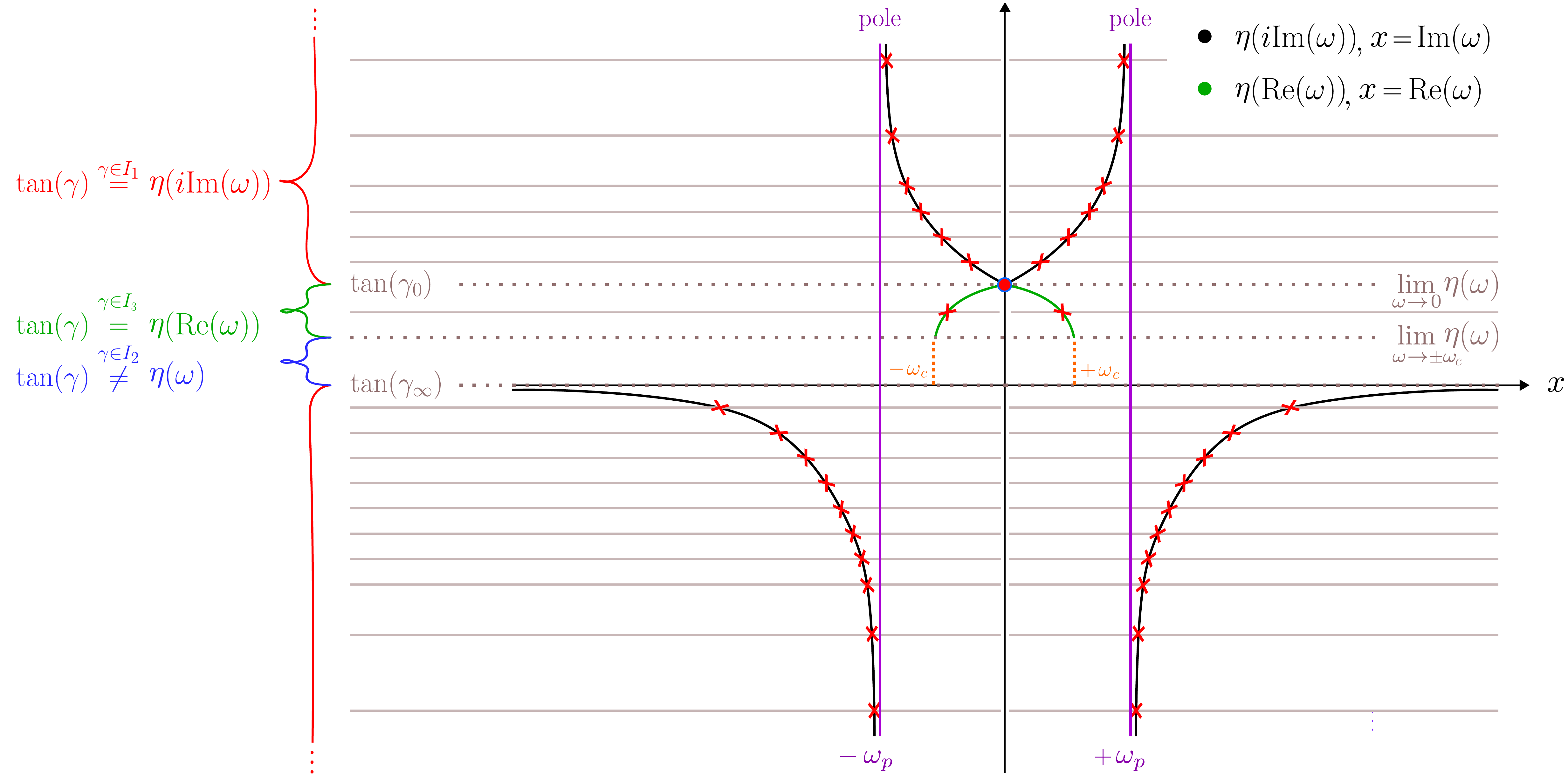}
\caption{The behavior of $\eta$ for frequencies such that $\omega^2\in\mathbb{R}$ and $\nu\in(1,2)$. Note that $|\omega_c|$ might be larger than $|\omega_p|$, i.e. in this illustration, the $x$-axis does not necessarily
have the same scale for $x=\Re(\omega)$ and $x=\Imag(\omega)$.}
\label{fig:kappa from 1 to 2 k+1}
\end{figure}

\noindent Analogously to the case $\kappa=-1$, and using the same notation, define
\begin{subequations}
\begin{align}
  &\zeta(\omega) := \tan(\gamma) - \eta(\omega),\label{eq:zeta = tan - kappa} \\
  &\eta(\omega) :=\frac{\Gamma(a+b-c)}{\Gamma(c-a-b)}\frac{\Gamma(1-a)\Gamma(c-a)}{\Gamma(b)\Gamma(b-c+1)}.\label{eq: kappa definition k +1}
\end{align}
\end{subequations}
\noindent Observe that $\eta(\omega)$ is real-valued whenever $\omega^2\in\mathbb{R}$, as illustrated in Figure \ref{fig:kappa on the complex plane k=+1}.
In addition,
\begin{subequations}
\begin{align}
  &\lim\limits_{\omega\rightarrow 0} \eta(\omega) =\frac{\Gamma (\nu )}{\Gamma (-\nu )} \frac{\Gamma \left(\frac{1-\nu}{2} +\frac{\sqrt{1-2 \lambda }}{2}\right)^2}{ \Gamma \left(\frac{1+\nu}{2}+\frac{ \sqrt{1-2 \lambda }}{2}\right)^2}, \label{eq: limits of Xi k=+1at omega 0}\\
  &\lim\limits_{\omega\rightarrow \omega_p} \eta(\omega) = \infty , \label{eq: limits of Xi k=+1 at omega omegap}\\
  &\lim\limits_{\omega\rightarrow \infty} \eta(\omega) = 0 ,\label{eq: limits of Xi k=+1 at omega infty}\\
  &\lim\limits_{\omega\rightarrow \omega_c} \eta(\omega) = \frac{\Gamma(\nu )}{\Gamma (-\nu )} \frac{\Gamma \left(\frac{1-\nu}{2} -i\frac{\sqrt{1-2 \lambda }}{2}\right)\Gamma \left(\frac{1-\nu}{2} +i\frac{\sqrt{1-2 \lambda }}{2}\right)}{\Gamma \left(\frac{1+\nu}{2} -i\frac{\sqrt{1-2 \lambda }}{2}\right)\Gamma \left(\frac{1+\nu}{2} +i\frac{\sqrt{1-2 \lambda }}{2}\right)} .\label{eq: limits of Xi k=+1 at omega_c}
\end{align}
\end{subequations}
\noindent Note that the first limit, $\eta(0)$, is negative for $\nu\in(0,1)$ and positive for $\nu\in(1,2)$.
The level sets of $\eta$ for $\kappa=+1$ are similar to the ones obtained in the previous two cases. However, in this case, the normalization has zeros not only for $\omega=i\Imag(\omega)$ with $\Imag(\omega)<0$, but also with $\Imag(\omega)>0$, and for real frequencies as well. For $\Imag(\omega)<0$, $\eta(i\Imag(\omega))$ behaves like the function $\Xi(i\Imag(\omega))$, as in Figure \ref{fig:Xi from 0 to 1} for $\nu\in(0,1)$, and as illustrated in Figure \ref{fig:Xi from 1 to 2} for $\nu\in(1,2)$.
For $\Imag(\omega)>0$, we have: $\eta(i\Imag(\omega))=\eta(-i\Imag(\omega))$. The zeros associated to the real frequencies, namely $\omega=\Real(\omega)\in\left(-\omega_c,\omega_c\right)$ for $\omega_c$ as in \eqref{eq: limiting value of real frequency k=+1}, yield an additional interval of boundary conditions for which there are bound states, given by
\begin{equation}
  \gamma\in \left( \arctan(\eta(\omega_c) + \pi), \arctan(\eta(0)+\pi )  \right).
\end{equation}
Figures \ref{fig:kappa from 0 to 1 k+1} and \ref{fig:kappa from 1 to 2 k+1} illustrate the behavior of $\eta(\omega)$, for $\nu\in(0,1)$ and $\nu\in(1,2)$, respectively, and including both curves for $\omega=i\Imag(\omega)$ and $\omega=\Real(\omega)$.

\vspace{1.5cm}
In conclusion, for $\nu\in(1,2)$, we can impose either Dirichlet boundary
conditions on all modes, or we can choose mode-dependent boundary conditions that respect the regime for which there are no bound states
\begin{equation}
  \label{eq:004urjf}
\gamma\in[0,\gamma^\lambda_{c}), \text{ with }\gamma^\lambda_{c} := \arctan\left( \eta(\omega_c)\right)\in(0,\pi).
\end{equation}
For $\nu\in(0,1)$, there are no bound states for $\gamma\in\left[0,\pi/2\right]$ and we can also impose the mode-dependent boundary conditions as given above by Equation \eqref{eq:004urjf}.
\section{Two-point functions for general boundary conditions}
\label{sec: Two-point functions for general boundary conditions}

 In this section, we address the question whether one can construct two-point functions of a free, scalar, massive quantum field theory on Lifshitz spacetime and on Lifshitz topological black holes for the admissible class of boundary condition of Robin type individuated in the previous sections. Inasmuch as they are static spacetimes, one can follow the same construction used \cite{Dappiaggi:2016fwc,Bussola:2017wki,Bussola:2018iqj,Dappiaggi:2017wvj,Dappiaggi:2018xvw,Dappiaggi:2018pju,Campos:2020lpt}. More specifically, the underlying global, timelike Killing field identifies a Hamiltonian and, accordingly, one can select an associated ground state for each admissible boundary condition. Using general properties of quantum field on curved backgrounds, one can infer a priori that such states are physically sensible, namely they satisfy the local Hadamard condition,
due to a general result of Sahlmann and Verch \cite{sahlmann2000passivity}. Here we do not enter into the details of such property, leaving an interested reader to the review \cite{Khavkine:2014mta} or to \cite{Kay:1988mu}.

 First, in Section \ref{sec: The construction}, we describe the framework
of the construction of the two-point functions, which apply to all spacetimes considered in this work. Then, we employ it explicitly on Lifshitz spacetime, on a hyperbolic and on a spherical Lifshitz topological black hole, respectively in Sections \ref{sec: Two-point functions on Lifshitz spacetime}, \ref{sec: Two-point functions on a hyperbolic Lifshitz black hole} and \ref{sec: Two-point functions on a spherical Lifshitz black hole}.
%
\subsection{The construction of physically-sensible two-point functions}
\label{sec: The construction}

On $\mathsf{Lif}_\kappa$, a two-point function of a free, scalar, massive
Klein-Gordon theory is a bidistribution $G^{+} \in \mathcal{D}'(\mathsf{Lif}_\kappa \times \mathsf{Lif}_\kappa)$ that is positive
 \begin{equation}
   \label{eq:positive}
 G^{+}(f,f) \geq 0, \forall \, f\in C_0^\infty(\mathsf{Lif}_\kappa)
 \end{equation}

 \noindent and solves the Klein-Gordon equation in each entry
 \begin{equation}
   \label{eq:bisolution}
 (P \otimes \mathbbm{1})G^{+}(f,f') = (\mathbbm{1} \otimes P)G^{+}(f,f') = 0,
 \end{equation}

\noindent for $P$ as in Equation \eqref{eq: KG equation general} and $ f,f'\in C_0^\infty(\mathsf{Lif}_\kappa)$.  If $E$ is the causal propagator, defined as the Green function of the Klein-Gordon equation, then the antisymmetric
part of $G^{+}$ satisfies
 \begin{equation}\label{eq:antisymmetric}
 iE(f, f') = G^{+}(f,f') - G^{+}(f',f).
 \end{equation}

To guarantee that the canonical commutation relations of the underlying quantum field theory hold, we impose the initial conditions ---see, for example \cite[Ch.3]{Kay:1988mu}--- at the level of integral kernel
\begin{subequations}
 \begin{align}
E(x,x')\mid_{t=t'} &=0, \label{eq:ccr1}\\
 \partial_t E(x,x')\mid_{t=t'} &= -\partial_{t'} E(x,x')\mid_{t=t'}
= \frac{\delta(\underline{x}-\underline{x}')}{q(x)},\label{eq:ccr2}
 \end{align}
\end{subequations}

\noindent where $x=(t,\underline{x})\in\mathsf{Lif}_\kappa$ and where $q$ is the function introduced in Equation \eqref{eq:measure mu(r)}. An analogous expression holds true for $x'$. Given that $\mathcal{M}$ admits a Killing field in the time direction, analogously to the ansatz for $\Psi$ \eqref{eq: Fourier expansion of scalar field}, we consider the ansatz for the
integral kernel of $G^{+}$
 \begin{equation}\label{eq: G+}
  G^{+}(x,x') = \lim_{\varepsilon \to 0^{+}}\int\limits_{\sigma(\triangle)}d\Sigma(\ell,m)\int_{0}^{\infty} d\omega e^{-i\omega(t-t'-i\varepsilon)}\widehat{G}_\omega(r,r')Y_\ell^m(\theta,\varphi)Y_\ell^m(\theta',\varphi'),
 \end{equation}

 \noindent where $i\varepsilon$ is a standard regularization, the limit is understood in the weak sense, $Y_\ell^m(\theta,\varphi)$ are the real-valued eigenfunctions of the Laplacian operator $\triangle$ with spectrum $\sigma(\triangle)$ and associated measure $d\Sigma(\ell,m)$, as mentioned in Section \ref{sec: Klein-Gordon Field}. For convenience, we take the harmonics normalized such that their completeness relation reads:

 \begin{equation}\label{eq: completeness relation harmonics}
\int\limits_{\sigma(\triangle)}d\Sigma(\ell,m)Y_\ell^m(\theta,\varphi)Y_\ell^m(\theta',\varphi')=\delta(\theta-\theta')\delta(\varphi-\varphi').
 \end{equation}

 \noindent With the ansatz \eqref{eq: G+}, the first initial condition, given by Equation \eqref{eq:ccr1}, is automatically satisfied if $\widehat{G}_\omega(r,r')=\widehat{G}_\omega(r',r)$. Furthermore, if $\widehat{G}_\omega(r,r')$ is also invariant under the mapping $\omega\mapsto -\omega$, and taking into account the completeness relations of the harmonics, then condition \eqref{eq:ccr2} yields
 \begin{equation}\label{potkhp90i49k}
 \int_{\mathbb{R}} d\omega \; \omega \widehat{G}_\omega(r,r') = \frac{\delta(r-r')}{q(r)}.
 \end{equation}

At the same time, for $L_{\omega^2}$ as the Sturm-Liuville operator defined in \eqref{eq: Sturm-Liouville operator L}, the radial Green function $\mathcal{G}_\omega$ is a solution of
 \begin{equation}\label{1gpr990b}
 ((L_{\omega^2}-\omega^2) \otimes \mathbbm{1})\mathcal{G}_\omega(r,r') =
(\mathbbm{1} \otimes (L_{\omega^2}-\omega^2))\mathcal{G}_\omega(r,r') =
\frac{\delta(r-r')}{q(r)},
 \end{equation}
with measure $q(r)$ defined in \eqref{eq:measure mu(r)}. In addition,
the spectral resolution of the radial Green function \cite{greenBook,Zettl:2005} reads
 \begin{equation}
   \label{eq:spectral resolution radial green function}
     \frac{1}{2\pi i}\oint_{\mathcal{C}^\infty} d(\omega^2) \mathcal{G}_\omega(r,r') + \frac{1}{2\pi i}\sum_{\text{poles}} \Res[\mathcal{G}_\omega(r,r')] = -\frac{\delta(r-r')}{s(r)} \text{,}
 \end{equation}

\noindent where $\mathcal{C}^{\infty}$ corresponds to a contour that, in the limit of infinite radius, covers the entire region on the $\omega^2$-complex plane where $\mathcal{G}_\omega$ is meromorphic. Note that when the radial Green function has no poles, the summation vanishes and, if we can reduce the contour integral to an integral over the real line, we can obtain $\widehat{G}_\omega(r,r')$ directly by comparing \eqref{potkhp90i49k} with \eqref{eq:spectral resolution radial green function}
\begin{equation}
  \label{eq:read off radial green from spectral resolution}
    \int_{\mathbb{R}} d\omega \, \omega \widehat{G}_\omega(r,r')=-\frac{1}{2\pi i}\oint_{\mathcal{C}^\infty} d(\omega^2) \mathcal{G}_\omega(r,r').
\end{equation}

Subsequently, for each spacetime $\mathsf{Lif}_\kappa$, we write the spectral resolution of the radial Green function, identify the appropriate contour, and show that it can be reduced to an integral over real frequencies. With that, we  extract a suitable definition of the only remaining unknown function $\widehat{G}_\omega(r,r')$. A two-point function as constructed above, of the form \eqref{eq: G+}, satisfies the canonical commutation relations and characterizes a ground-state of local Hadamard form \cite{sahlmann2000passivity}. With such ground-state in hands, a thermal state can be directly constructed by replacing the time-dependent function. That is, if the two-point function of the ground-state \eqref{eq: G+} is well-defined, then a KMS state at inverse temperature $\beta$ with respect to $\partial_t$ is characterized by two-point function of the form
\begin{subequations}
\begin{equation}\label{eq: G+ KMS}
 G^{+}_\beta(x,x') = \lim_{\varepsilon \to 0^{+}}\int\limits_{\sigma(\triangle)}d\Sigma(\ell,m)\int_{0}^{\infty} d\omega  T_\omega(t,t')\widehat{G}_\omega(r,r')Y_\ell^m(\theta,\varphi)Y_\ell^m(\theta',\varphi'),
\end{equation}
with
\begin{equation}
  \label{eq:T(t) KMS time part}
  T_\omega(t,t'):=\frac{ e^{\beta\omega}e^{-i \omega (t-t'-i\varepsilon)}+e^{+i \omega (t-t'+i\varepsilon)} }{e^{\beta\omega}-1}.
\end{equation}
\end{subequations}

 \subsection{Two-point functions on Lifshitz spacetime $\mathsf{Lif}_0$}
 \label{sec: Two-point functions on Lifshitz spacetime}

\begin{figure}[H]
  \centering
   \includegraphics[align=c,width=.4\textwidth]{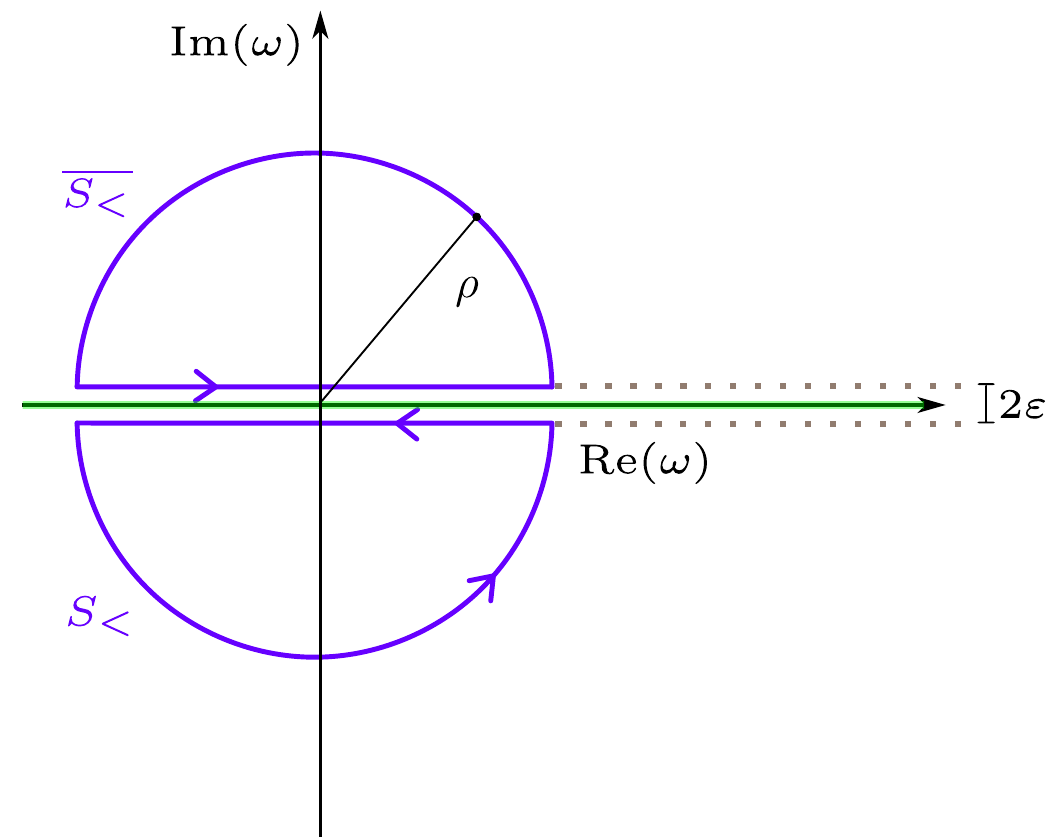}
  \caption{Contour of two semi-circles on the $\omega$-complex plane.}
  \label{fig:contour 2 semi circles}
    \end{figure}

Bearing in mind the analysis of Section \ref{sec: the radial equation equation on Lifshitz spacetime, k=0}, for an effective mass such that $\nu\in(0,1)\cup(1,2)$, and for boundary conditions parametrized by $\gamma$ at radial infinity, the radial Green function on $\mathsf{Lif}_0$, given by Equation \eqref{eq: green function k 0}, reads
\begin{align*}
\mathcal{G}_{\omega}(r,r') =\frac{1}{ \mathcal{N}_{\omega}}\left( \Theta(r'-r) R_{0}(r)R_{\gamma}(r') + \Theta(r-r')R_{0}(r')R_{\gamma}(r)\right).
\end{align*}
Recalling that $ R_{0}(r)$ is defined by parts, we define $  \mathcal{G}_{\omega}^{\mini{<}}(r,r')$ and $  \mathcal{G}_{\omega}^{\mini{>}}(r,r')$ implicitly by
\begin{align}
\mathcal{G}_{\omega}(r,r')= \Theta(-\Imag(\omega))\mathcal{G}_{\omega}^<(r,r')+\Theta(\Imag(\omega))\mathcal{G}_{\omega}^>(r,r').
\end{align}
In view of the symmetry properties described in Section \ref{sec: Symmetries under conjugation and reflection k=0},  it holds
\begin{align}
  \label{eq: conjugate G omega < = G > conjugate omega}
\overline{\mathcal{G}_{\omega}^{\mini{<}}(r,r')}=\mathcal{G}_{\overline{\omega}}^{\mini{>}}(r,r').
\end{align}
For $\nu\in(0,1)$ and $\gamma\in[0,\pi/2]$, the radial Green function has
no poles, but it is not defined for frequencies with $\Imag(\omega)=0$.
Therefore, on the $\omega$-complex plane, the suitable contour of the spectral resolution \eqref{eq:spectral resolution radial green function} consists of two semi-circles, as in Figure \ref{fig:contour 2 semi circles}.
Taking into account Equation \eqref{eq: conjugate G omega < = G > conjugate
omega}, we write
\begin{align}
  \label{eq:oirjfoijrf}
  \oint_{\mathcal{C}^\infty} d\omega \,\omega \mathcal{G}_{\omega}(r,r')=\lim\limits_{\varepsilon\rightarrow0}\lim\limits_{\rho\rightarrow\infty} \left\{\oint_{S_{\mini{<}}} d\omega\omega \mathcal{G}_{\omega}^{\mini{<}}(r,r') +\oint_{\overline{S_{\mini{<}}}} d\omega \omega\overline{\mathcal{G}_{\overline{\omega}}^{\mini{<}}(r,r')}\right\}.
\end{align}
Each contour integral of the expression above has two components given by
the right and left propagating terms of $\mathcal{G}_\omega$. For $\Imag(\omega)<0$, we define the right component as
\begin{align}
 \mathcal{G}_{\omega}^{\mini{<,R}}(r,r') = \frac{1}{ \mathcal{N}_{\omega}} \Theta(r'-r) R_{1(0)}(r)R_{\gamma}(r'),
\end{align}
while $\mathcal{G}_{\omega}^{\mini{<,L}}(r,r') =\mathcal{G}_{\omega}^{\mini{<}}(r,r')-\mathcal{G}_{\omega}^{\mini{<,R}}(r,r')  $.
Therefore, the right-hand side of Equation \eqref{eq:oirjfoijrf} can be decomposed as
\begin{align}
  \label{eq:oirjfoijrf2}
\lim\limits_{\varepsilon\rightarrow0}\lim\limits_{\rho\rightarrow\infty} \left\{\oint_{S_{\mini{<}}} d\omega\omega\left[\mathcal{G}_{\omega}^{\mini{<,R}}(r,r') + \mathcal{G}_{\omega}^{\mini{<,L}}(r,r')\right] +\oint_{\overline{S_{\mini{<}}}} d\omega \omega \left[\overline{\mathcal{G}_{\overline{\omega}}^{\mini{<,R}}(r,r')}+\overline{\mathcal{G}_{\overline{\omega}}^{\mini{<,L}}(r,r')}\right]\right\}.
\end{align}

\noindent First, let us focus on the integration over the right components. That is, consider
\begin{align}
I^{\mini{R}}:=\oint_{S_{\mini{<}}} d\omega\omega \mathcal{G}_{\omega}^{\mini{<,R}}(r,r') +\oint_{\overline{S_{\mini{<}}}} d\omega \omega \overline{\mathcal{G}_{\overline{\omega}}^{\mini{<,R}}(r,r')}.
\end{align}

\noindent Since the integrand has no poles within the contour, by the same argument
of \cite[Appx.A]{Bussola:2017wki}, or by a direct application of Jordan's
Lemma, the arc integrals vanish. Thus, substituting the definitions and recalling that the radial solutions satisfy properties \eqref{eq: properties basis R(r) at 0 k=0}, it follows
\begin{align}
  \label{eq:weiopjwopkefop}
\lim\limits_{\varepsilon\rightarrow0}\lim\limits_{\rho\rightarrow\infty} I^{\mini{R}} = \Theta(r'-r)  \int_{\mathbb{R}}d\tilde{\rho}\,\tilde{\rho} \left[ \left(\frac{1}{ \mathcal{N}_{\omega}}  R_{1(0)}(r) - \frac{1}{\overline{ \mathcal{N}_{ \omega}}}   \overline{R_{1(0)}(r)}\right) R_{\gamma}(r')\right]\Bigg|_{\omega=\tilde{\rho}}.
\end{align}
Using the expressions \eqref{eq: fundamental relation R10 as R1infty R2infty k=0}, \eqref{eq: normalization k = 0 omega A0 and B0} and \eqref{eq: property normalization and green omega conjugate k=0}, we simplify the term within the brackets in Equation \eqref{eq:weiopjwopkefop} and the contribution to the contour integral coming from the right propagating terms is

\begin{align}
  \label{eq:weiopjwopkefop111}
\lim\limits_{\varepsilon\rightarrow0}\lim\limits_{\rho\rightarrow\infty} I^{\mini{R}} = \Theta(r'-r)  \int_{\mathbb{R}}d\tilde{\rho}\,\tilde{\rho}\left[\frac{1}{4\nu}\frac{A_0\overline{B_0}-B_0\overline{A_0}}{|B_0\cos(\gamma)-A_0\sin(\gamma)|^2} R_{\gamma}(r) R_{\gamma}(r')\right]\Bigg|_{\omega=\tilde{\rho}}.
\end{align}
We can obtain the contribution from the left propagating terms by an analogous computation. However, noting that the integrand above is invariant under the mapping $r\leftrightarrow r'$, we can see that it is not necessary to compute it, and that adding both contributions is equivalent to scratching out the Heaviside function of the right-hand side of \eqref{eq:weiopjwopkefop111}.

All in all, we obtain
\begin{align}
  \label{eq:weiopjwopkefop111222}
\oint_{\mathcal{C}^\infty} d\omega \,\omega \mathcal{G}_{\omega}(r,r') =  \int_{\mathbb{R}}d\omega\,\omega\left[ \frac{1}{4\nu}\frac{A_0\overline{B_0}-B_0\overline{A_0}}{|B_0\cos(\gamma)-A_0\sin(\gamma)|^2} R_{\gamma}(r) R_{\gamma}(r')\right],
\end{align}
which yields, by comparing it with Equation \eqref{eq:read off radial green from spectral resolution}, the function:
\begin{equation}
  \label{eq:G+ hat k=0}
    \widehat{G}_\omega(r,r')=\frac{1}{4 \pi \nu}\frac{\Imag{(B_0\overline{A_0})}}{|B_0\cos(\gamma)-A_0\sin(\gamma)|^2} R_{\gamma}(r) R_{\gamma}(r').
\end{equation}
Note that $\widehat{G}_\omega(r,r')$ is, as needed to guarantee the canonical commutation relations, invariant under both transformations $\omega\leftrightarrow-\omega$ and $r\leftrightarrow r'$.

Let $Y_\ell^m(\theta,\varphi)$ be the real-valued eigenfunctions of the Laplacian operator on the two-dimensional Euclidean space. A two-point function with integral kernel

\begin{equation}\label{eq: G+ k=0}
 G^{+}(x,x') = \lim_{\varepsilon \to 0^{+}}\int_{\mathbb{R}}d\ell\int_{\mathbb{R}}d m\int_{0}^{\infty} d\omega e^{-i\omega(t-t'-i\varepsilon)}\widehat{G}_\omega(r,r')Y_\ell^m(\theta,\varphi)Y_\ell^m(\theta',\varphi'),
\end{equation}

characterizes a ground state for the free, scalar, massive Klein-Gordon field on a Lifshitz spacetime. By the same token, a two-point of the form

\begin{equation}\label{eq: G+ k=0 KMS}
 G_\beta^{+}(x,x') = \lim_{\varepsilon \to 0^{+}}\int_{\mathbb{R}}d\ell\int_{\mathbb{R}}dm\int_{0}^{\infty} d\omega T_\omega(t,t')\widehat{G}_\omega(r,r')Y_\ell^m(\theta,\varphi)Y_\ell^m(\theta',\varphi'),
\end{equation}

with $T_\omega(t,t')$ given by \eqref{eq:T(t) KMS time part}, characterizes a KMS state at inverse-temperature $\beta$ with respect to the Killing
field $\partial_t$ for the free, scalar, massive Klein-Gordon field on a Lifshitz spacetime. Both two-point functions are of local Hadamard form and satisfy the canonical commutation relations.

\subsection{Two-point functions on a hyperbolic Lifshitz black hole $\mathsf{Lif}_{-1}$}
\label{sec: Two-point functions on a hyperbolic Lifshitz black hole}

On the hyperbolic Lifshitz black hole, $\mathsf{Lif}_{-1}$, the computation regarding the spectral decomposition of the radial Green function is equivalent to the one described in the previous section, for the $\kappa=0$ case. The radial Green function is also defined by parts with respect to the sign of the imaginary part of the frequencies, the suitable contour is also given by the one in Figure \ref{fig:contour 2 semi circles} and
the steps to follow are the same. Invoking the results of Section \ref{sec: The radial equation on the hyperbolic Lifshitz black hole, k-1} regarding the Klein-Gordon equation, we consider values of mass for which no bound states emerge, as given in \eqref{eq: no poles km1 neumann}, and the acceptable boundary conditions, as described in Section \ref{sec: on the existence of bound states for k=-1}. Analogously to \eqref{eq:G+ hat k=0}, we obtain
\begin{equation}
  \label{eq:G+ hat k=-1}
    \widehat{G}_\omega(r,r')=\frac{1}{ \pi L^4 \nu}\frac{\Imag{(B_{\mini{-1}}\overline{A_{\mini{-1} } })} }{|B_{\mini{-1}}\cos(\gamma)-A_{\mini{-1}}\sin(\gamma)|^2} R_{\gamma}(r) R_{\gamma}(r').
\end{equation}
With $  \widehat{G}_\omega(r,r')$ as above, $T_\omega(t,t')$ as in \eqref{eq:T(t) KMS time part} and letting $Y_\ell^m(\theta,\varphi)$ be the real-valued eigenfunctions of the Laplacian operator on the $2$-dimensional hyperbolic space, it follows that
\begin{equation}\label{eq: G+ k=0}
 G^{+}(x,x') = \lim_{\varepsilon \to 0^{+}} \int_{0}^\infty d \ell\sum_{m=0}^\infty\int_{0}^{\infty} d\omega e^{-i\omega(t-t'-i\varepsilon)}\widehat{G}_\omega(r,r')Y_\ell^m(\theta,\varphi)Y_\ell^m(\theta',\varphi'),
\end{equation}
and
\begin{equation}\label{eq: G+ k=0 KMS}
 G_\beta^{+}(x,x') = \lim_{\varepsilon \to 0^{+}}\int_{0}^\infty d\ell\sum_{m=0}^\infty \int_{0}^{\infty} d\omega T_\omega(t,t')\widehat{G}_\omega(r,r')Y_\ell^m(\theta,\varphi)Y_\ell^m(\theta',\varphi'),
\end{equation}
are two-point functions that characterize, respectively, a ground state and a KMS state at inverse-temperature $\beta$ with respect to the Killing
field $\partial_t$, for the free, scalar, massive Klein-Gordon field on a
hyperbolic Lifshitz black hole, which are of local Hadamard form and satisfy the canonical commutation relations.

\subsection{Two-point functions on a spherical Lifshitz black hole $\mathsf{Lif}_1$}
\label{sec: Two-point functions on a spherical Lifshitz black hole}

The spherical Lifshitz black hole constitutes a manifestly different scenario from the previous two cases. Recall that in this case the radial Green function is defined for all frequencies except for the real-valued ones with $|\omega|>\omega_c$, as stated in Equation \eqref{eq: real upsilon
= 0 omega interval}, which is tantamount to assume $\Real{\Upsilon}>0$ together with $\xi<\frac{1}{2}$.
We consider the regime of boundary conditions for which there are no bound states, namely for $\nu\in(0,1)$ we take $\gamma\in[0,\pi/2]$, and for $\nu\in(1,2)$, we assume either $\gamma=0$ or $\gamma$ is mode-dependent, as detailed in Section \ref{sec: On the existence of bound states for k=+1}.

\begin{figure}[H]
   \centering
    \includegraphics[align=c,width=.4\textwidth]{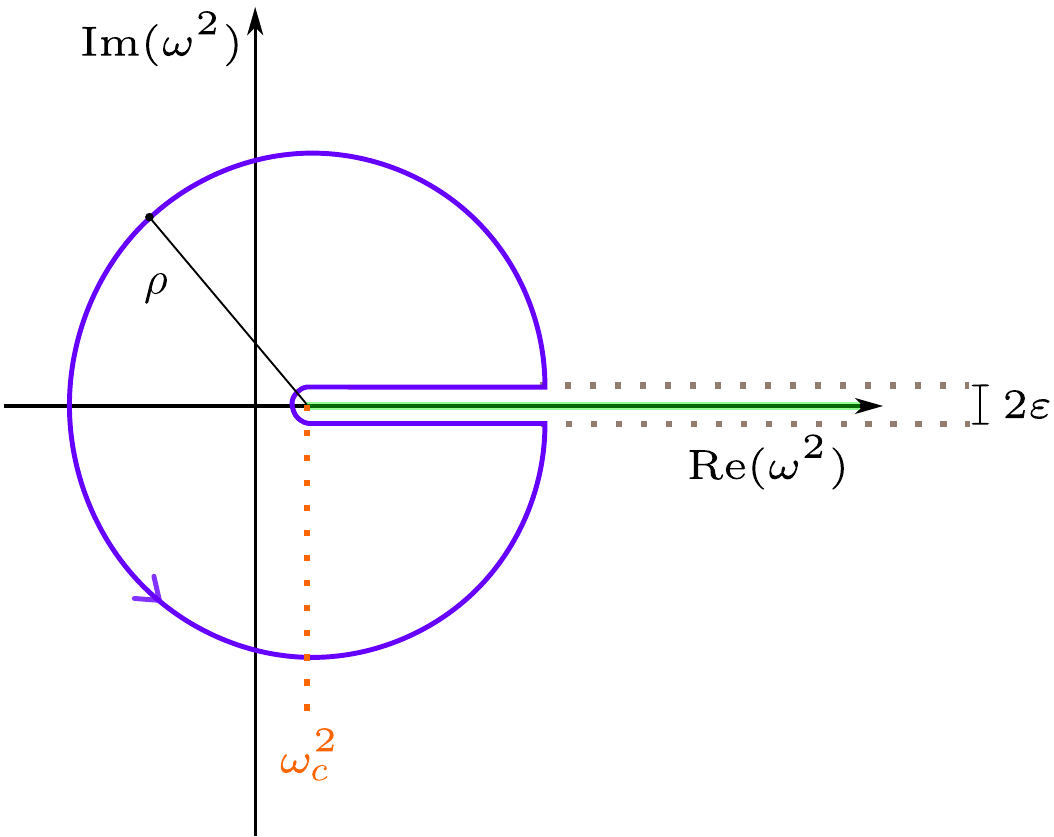}
   \caption{``Pac-man'' contour on the $\omega^2$-complex plane.}
   \label{fig:contour pacman}
     \end{figure}

For $\Real{\Upsilon}>0$, the radial Green function, given by Equation \eqref{eq: green function k +1} with $R_{2(0)}(r)$ determined by \eqref{eq: basis R(s) k +1}, reads:
\begin{align}
\mathcal{G}_{\omega}(r,r') =\frac{1}{ \mathcal{N}_\omega}\left( \Theta(r'-r) R_{2(0)}(r)R_{\gamma}(r') + \Theta(r-r')R_{2(0)}(r')R_{\gamma}(r)\right).
\end{align}
Let us define the right propagating term by
\begin{align}
\mathcal{G}_{\omega}^{\mini{R}}(r,r') =\frac{1}{ \mathcal{N}_\omega} \Theta(r'-r) R_{2(0)}(r)R_{\gamma}(r')
\end{align}
and the left propagating term by $\mathcal{G}_{\omega}^{\mini{L}}(r,r')=\mathcal{G}_{\omega}(r,r')-\mathcal{G}_{\omega}^{\mini{R}}(r,r')$. The suitable contour for the spectral resolution is given by the ``pac-man'' contour on the $\omega^2$-complex plane as illustrated in Figure \ref{fig:contour pacman}. As a consequence
\begin{align}
  \label{eq:oirjfoijrf}
  \oint_{\mathcal{C}^\infty} d(\omega^2) \mathcal{G}_{\omega}(r,r')=\lim\limits_{\varepsilon\rightarrow0}\lim\limits_{\rho\rightarrow\infty} \oint\limits_{\mini{\text{``pac-man''}}} d(\omega^2)\left\{\mathcal{G}_{\omega}^{R}(r,r') + \mathcal{G}_{\omega}^{L}(r,r')\right\}.
\end{align}
\noindent First, we note that the arc integrals vanish, as in the previous two cases. The integral over $\rho$ of the right propagating terms, denoted by $ I^{\mini{R}}$, yields
\begin{align}
  \label{eq:eofpoekfoe}
   \lim\limits_{\varepsilon\rightarrow0}\lim\limits_{\rho\rightarrow\infty} I^{\mini{R}}= & \Theta(r'-r)\left\{ \int_{+\omega_c}^{+\infty}  d\tilde{\rho}\, \left(\frac{1}{ \mathcal{N}_\omega} R_{2(0)}(r) - \frac{1}{ \mathcal{N}_{\overline{\omega}}}  \overline{R_{2(0)}(r)}\right)  R_{\gamma}(r') \right\}\Bigg|_{\omega=\sqrt{\tilde{\rho}}}.
\end{align}

\noindent Defining $\widetilde{\omega} = \omega - \omega_c$, which depends on $\lambda$ because $\omega_c$ does, and invoking the same arguments given in the previous two spacetimes, we obtain the function $\widehat{G}_{\widetilde{\omega}}(r,r')$ for the spherical case
\begin{equation}
  \label{eq:G+ hat k=0}
    \widehat{G}_{\widetilde{\omega}}(r,r')=\left[\frac{1}{ \pi L^4 \nu}\frac{\Imag{(B_{\mini{+1}}\overline{A_{\mini{+1} } })} }{|B_{\mini{+1}}\cos(\gamma)-A_{\mini{+1}}\sin(\gamma)|^2} R_{\gamma}(r) R_{\gamma}(r')\right]\Bigg|_{\omega=\widetilde{\omega} + \omega_c}
\end{equation}

Last, let $Y_\ell^m(\theta,\varphi)$ be the spherical harmonics and $T_{\widetilde{\omega}}(t,t')$ as in \eqref{eq:T(t) KMS time part}. Two-point functions with integral kernels given by
\begin{equation}\label{eq: G+ k=0}
 G^{+}(x,x') = \lim_{\varepsilon \to 0^{+}}\sum_{\ell=0}^\infty\sum_{m=-\ell}^\ell\int_{0}^{\infty} d\widetilde{\omega} e^{-i\widetilde{\omega}(t-t'-i\varepsilon)}\widehat{G}_{\widetilde{\omega}}(r,r')Y_\ell^m(\theta,\varphi)Y_\ell^m(\theta',\varphi'),
\end{equation}
and
\begin{equation}
  \label{eq: G+ k=0 KMS}
 G_\beta^{+}(x,x') = \lim_{\varepsilon \to 0^{+}}\sum_{\ell=0}^\infty\sum_{m=-\ell}^\ell\int_{0}^{\infty} d\widetilde{\omega} T_{\widetilde{\omega}}(t,t')\widehat{G}_{\widetilde{\omega}}(r,r')Y_\ell^m(\theta,\varphi)Y_\ell^m(\theta',\varphi'),
\end{equation}
characterize, respectively, a ground and a KMS state at inverse-temperature $\beta$ with respect to the Killing field $\partial_t$ for the free, scalar, massive Klein-Gordon field on $\mathsf{Lif}_1$. Recall that, in addition, they are of local Hadamard form and they satisfy the canonical commutation relations per construction.

\section{Conclusion}

We showed that a free, scalar field with effective mass $\mu^2\in \left(-\frac{4}{L^2},-\frac{3}{L^2}\right)$ on a class of Lifshitz type spacetime, $\mathsf{Lif}_\kappa$, can be endowed with a class of Robin boundary conditions ruled by a continuous parameter $\gamma\in[0,\pi/2]$. Each of these boundary conditions yield non-equivalent dynamics and different two-point functions, thus generalizing the standard quantization procedure which relies on Dirichlet boundary conditions. From a physical viewpoint all these additional options appear to be a legitimate choice since, on the
one hand, we are still describing a closed system, while, on the other hand, we can construct an associated two-point function, locally of Hadamard form. Hence, physical observables, such as the renormalized stress-energy tensor, can be constructed yielding finite expectation values and finite quantum fluctuations.

It is worth mentioning that one might foresee to consider a more general class of boundary conditions, much in the spirit of what happens on asymptotically AdS spacetimes, see {\em e.g.} \cite{Dappiaggi:2018pju, Dappiaggi:2021wtr}. In particular, for the range of effective mass $\mu^2\in \left(-\frac{4}{L^2},-\frac{3}{L^2}\right)\cup\left(-\frac{3}{L^2},-\frac{2}{L^2}\right)$, $\lambda$-mode dependent boundary conditions are admissible. Similar options have been recently investigated in the context of AdS spacetimes \cite{Barroso:2019cwp}. Hence an analysis in this direction would be certainly an interesting research project, which we hope to undertake in the next future.

In addition we emphasize that we expect that the framework considered in this paper can be generalized to higher dimensional Lifshitz spacetimes of critical exponent $z=2$. For other values of $z>2$ a case-by-case study is necessary, since the classification, and hence the properties of the solutions of the radial part of the Klein-Gordon equation are strongly dependent on the choice of $z$.
Moreover, starting from this work, one can conceive other two additional research avenues. On the one hand one can perform a numerical analyses to
understand how the choice of boundary conditions and of topology affects vacuum fluctuations, much in the same spirit of the work presented in \cite{Morley:2020zcd} for topological black holes of Einstein gravity. On the other hand, one can compute the transition rate of an Unruh-DeWitt detector in a $\mathsf{Lif}_\kappa$ spacetime, following the same prescription of \cite{Campos:2020lpt} unveiling in particular how such detector distinguishes among the three Lifshitz spacetimes considered, or how it interacts differently with a massless hyperbolic black hole or with a Lifshitz
topological black holes.

\section{Acknowledgments}


The work of L.S.C is supported by a PhD scholarship of the University of Pavia, which is gratefully acknowledged. Part of this work is based on the MSc thesis of D.S. submitted for evaluation to the University of Pavia on the 26th of March 2021.

	\end{document}